\documentclass[12pt,dvips]{article}
\usepackage{rotating}
\usepackage{epsfig}
\usepackage{color}
\usepackage{cite}

\newcommand{\imag}{\Im {\rm m}}
\newcommand{\real}{\Re {\rm e}}
\newcommand{\tanb}{\tan \! \beta}

\newcommand{\mstauo}{m^2_{\tilde{\tau}_1}}
\newcommand{\mstaut}{m^2_{\tilde{\tau}_2}}
\newcommand{\ghat}{\hat{g}^2}

\newcommand{\htau}{\left| h_\tau \right|^2}

\newcommand{\bra}[1]{\langle #1|}
\newcommand{\ket}[1]{|#1\rangle}

\begin{document}

\def\thefootnote{\fnsymbol{footnote}}

{\small
\begin{flushright}
FERMILAB-PUB-12-466-T,
ANL-HEP-PR-12-58,
EFI-12-20, MAN/HEP/2012/11\\ 
KCL-PH-TH/2012-30, LCTS/2012-15, CERN-PH-TH/2012-195,
CNU-HEP-12-01 
\\
\end{flushright} }

\medskip

\begin{center}
{\bf 
{\LARGE {\color{red}CP}{\color{blue}super}{\color{green}H}{\color{black}2.3}:}\\
{\Large an Updated Tool for Phenomenology in the MSSM\\[2mm] 
with Explicit CP Violation} 
}
\end{center}

\smallskip

\begin{center}{\large
J.~S.~Lee$^{a,b}$,
M.~Carena$^c$,
J.~Ellis$^{d,e}$,
A.~Pilaftsis$^f$
and C.~E.~M.~Wagner$^{g,h}$}
\end{center}

\begin{center}
{\em $^a$ Department of Physics, Chonnam National University, \\
300 Yongbong-dong, Buk-gu, Gwangju, 500-757, Republic of Korea}\\[0.2cm]
{\em $^b$ Department of Physics, National Tsing Hua University, Hsinchu, Taiwan
300} \\[0.2cm]
{\em $^c$Fermilab, P.O. Box 500, Batavia IL 60510, U.S.A.}\\[0.2cm]
{\em $^d$Theoretical Particle Physics and Cosmology Group, Department of
  Physics, King's~College~London, London WC2R 2LS, United Kingdom}\\[0.2cm]
{\em $^e$Theory Division, CERN, CH-1211 Geneva 23, Switzerland}\\[0.2cm]
{\em $^f$Consortium for Fundamental Physics, School of Physics and Astronomy}\\
{\em University of Manchester, Manchester M13 9PL, United Kingdom}\\[0.2cm]
{\em $^g$HEP Division, Argonne National Laboratory,
9700 Cass Ave., Argonne, IL 60439, USA}\\[0.2cm]
{\em $^h$Enrico Fermi Institute, Univ. of Chicago, 5640
Ellis Ave., Chicago, IL 60637, USA}
\end{center}

\bigskip

\centerline{\bf ABSTRACT}
{\small
\medskip\noindent  We  describe the  Fortran  code {\tt  CPsuperH2.3},
which incorporates the following updates compared with its predecessor
{\tt  CPsuperH2.0}.   It   implements  improved  calculations  of  the
Higgs-boson masses and mixing  including stau contributions and finite
threshold effects on the  tau-lepton Yukawa coupling.  It incorporates
the LEP limits on  the processes $e^+ e^- \to H_i Z,  H_i H_j$ and the
CMS limits on  $H_i \to {\bar \tau} \tau$  obtained from 4.6~fb$^{-1}$
of data  at a  centre-of-mass energy of  7~TeV.  It also  includes the
decay mode  $H_i \to Z\gamma$  and the Schiff-moment  contributions to
the electric  dipole moments of  Mercury and Radium~225,  with several
calculational options  for the case of Mercury.   These additions make
{\tt CPsuperH2.3} a suitable  tool for analyzing possible CP-violating
effects in the MSSM in the era  of the LHC and a new generation of EDM
experiments}~\footnote{The   program  may   be   obtained  from   {\tt
    http://www.hep.man.ac.uk/u/jslee/CPsuperH.html},  or by contacting
  the first author at {\tt jslee@jnu.ac.kr}.}.

\vspace{0.2in}

\newpage

{\bf UPDATED VERSION PROGRAM SUMMARY}

\begin{small}
\noindent
{\em Manuscript Title:} CPsuperH2.3: 
an Updated Tool for Phenomenology in the MSSM
with Explicit CP Violation \\
{\em Authors:} J.S. Lee, M. Carena, J. Ellis, A. Pilaftsis and C.E.M. Wagner \\
{\em Program Title:} CPsuperH2.3 \\
{\em Journal Reference:}                                      \\
{\em Catalogue identifier:} ADSR\_v3\_0                                  \\
{\em Licensing provisions:} none                                   \\
{\em Programming language:} Fortran77  \\
{\em Computer:} PC running under Linux and computers in Unix environment \\
{\em Operating system:} Linux                                       \\
{\em RAM:} 32 Mbytes                                              \\
{\em Number of processors used:}                              \\
{\em Supplementary material:}                                 
{\em Classification:} 11.1 General, High Energy Physics and Computing\\
{\em External routines/libraries:}                                      \\
{\em Subprograms used:}                                       \\
{\em Catalogue identifier of previous version:} ADSR\_v2\_0               \\
{\em Journal reference of previous version:} Comput. Phys. Commun. 180(2009)312
\\
{\em Does the new version supersede the previous version?:} Yes   \\
{\em Nature of problem:}
The calculations of mass spectrum, decay  widths and branching ratios
of  the neutral and charged Higgs bosons in the Minimal Supersymmetric Standard
Model with explicit  CP violation have been improved.  The  program   is   based
on
renormalization-group-improved diagrammatic  calculations that include
dominant   higher-order   logarithmic   and  threshold   corrections,
$b$-quark  and $\tau$-lepton
Yukawa-coupling   resummation   effects  and   improved treatment
of Higgs-boson pole-mass shifts.
The couplings of the Higgs bosons to the Standard Model gauge bosons and fermions,
to their supersymmetric partners and all the trilinear and quartic Higgs-boson
self-couplings
are also calculated.
Also included are a full treatment of the $4 \times 4$ ($2\times 2$)
neutral (charged) Higgs propagator matrix together with the
 center-of-mass  dependent Higgs-boson  couplings to
gluons  and  photons,
and an integrated treatment of several $B$-meson
observables.
The new implementations include  
the EDMs of  Thallium, neutron,
Mercury, Deuteron,  Radium, and muon, as well as the  anomalous magnetic moment of
muon, $(g_\mu-2)$, the top-quark decays,
improved  calculations  of  the
Higgs-boson masses and mixing  including stau contributions,
the LEP limits, and the CMS limits on  $H_i \to {\bar \tau} \tau$.
It also  implements the decay mode  $H_i \to Z\gamma$  
and includes the corresponding Standard Model branching ratios of
the three neutral Higgs bosons in the array
{\tt GAMBRN(IM,IWB=2,IH)}.
\\
   \\
{\em Solution method:} One-dimensional numerical integration for several
Higgs-decay modes and EDMs,
iterative treatment of the threshold corrections and Higgs-boson pole masses,
and the numerical diagonalization of the neutralino mass matrix.\\
   \\
{\em Reasons for the new version:} Mainly to provide the full calculations of
the EDMs of  Thallium, neutron,
Mercury, Deuteron,  Radium, and muon as well as $(g_\mu-2)$,
improved  calculations  of  the
Higgs-boson masses and mixing  including stau contributions,
the LEP limits, the CMS limits on  $H_i \to {\bar \tau} \tau$,
the top-quark decays,
$H_i \to Z\gamma$  decay,
and the corresponding Standard Model branching ratios of
the three neutral Higgs bosons.
\\
   \\
{\em Summary of revisions:} Full calculations of
the EDMs of  Thallium, neutron,
Mercury, Deuteron,  Radium, and muon as well as $(g_\mu-2)$.
Improved treatment of Higgs-boson masses and mixing  including stau
contributions. The LEP limits. The CMS limits on  $H_i \to {\bar \tau} \tau$.
The top-quark decays.
The $H_i \to Z\gamma$  decay. The corresponding Standard Model branching ratios of
the three neutral Higgs bosons.
\\
   \\
{\em Restrictions:} No\\
   \\
{\em Unusual features:} No\\
   \\
{\em Additional comments:} No\\
   \\
{\em Running time:} Less than 1.0 second.\\
   \\

\end{small}

\newpage

\newpage
\section{Introduction}
\label{sec:intro}

Supersymmetry  is one of  the most  attractive possible  scenarios for
physics beyond  the Standard  Model in the  TeV energy range,  and the
minimal supersymmetric  extension of the Standard Model  (MSSM) is the
simplest framework  that incorporates  this scenario. The  MSSM allows
for  many possible CP-violating  phases, notably  those in  the SU(3),
SU(2) and  U(1) gaugino masses  $M_{3,2,1}$ and in the  trilinear soft
supersymmetry-breaking parameters associated with the third-generation
Yukawa couplings $A_{t,b,\tau}$.

{\tt  CPsuperH}~\cite{Lee:2007gn,Lee:2003nta} is an  evolving software
tool that  incorporates the effects of these  CP-violating phases into
the   calculation  of   Higgs  boson   masses,  couplings   and  other
properties. The previous version, {\tt CPsuperH2.0}~\cite{Lee:2007gn},
also incorporated  a number of $B$-physics  observables, including the
branching ratios for  $B_s \to \mu^+ \mu^-$, $B_d  \to \tau^+ \tau^-$,
$B_u \to \tau \nu$, $B  \to X_s \gamma$ and the CP-violating asymmetry
in  the  latter  decay,   $A_{CP}$,  as  well  as  the  supersymmetric
contributions   to   the   $B^0_{s,d}   -   {\bar   B^0}_{s,d}$   mass
differences. Also  included in {\tt CPsuperH2.0}  were calculations of
two-loop  supersymmetric Higgs  contributions to  the  electric dipole
moments  (EDMs) of  Thallium, the  electron and  muon~\footnote{For an
alternative code treating similar physics, see~\cite{FeynHiggs}.}.

After  the  recent  discovery  at  the LHC~\cite{ATLASCMS}  of  a  new
particle resembling  a Standard-Model-like Higgs  boson, Higgs physics
and searches  for associated CP-violating  effects are entering  a new
era, in  which the LHC experiments  are now also  probing directly the
possible  existences  of  heavier  supersymmetric  Higgs  bosons.   In
parallel, the LHCb experiment  is taking experiments on $B$-physics to
a new  level of precision,  and a new  round of experiments is  set to
improve  significantly  sensitivities  to   a  wider  range  of  EDMs,
including also Mercury and Radium. This juncture in the exploration of
physics at the TeV scale, both direct and indirect, is the appropriate
moment to document the capabilities of a new update of {\tt CPsuperH}.

The  followings  are  the  main  new  features  incorporated  in  {\tt
  CPsuperH2.3}.  As  described in  Section~2, it features  an improved
treatment of  CP-violating effects  on Higgs-boson masses  and mixing,
with  stau contributions  included as  in  Ref.~\cite{Choi:2000wz}. We
also take  consistently into account  finite radiative effects  on the
tau-lepton   Yukawa  coupling.   As   described  in   Section~3,  {\tt
  CPsuperH2.3} also  features an implementation  of the LEP  limits on
the processes  $e^+ e^- \to  H_i Z, H_i  H_j$~\cite{Schael:2006cr}, as
well as the limits on $H_i  \to {\bar \tau} \tau$ obtained by CMS with
4.6~fb$^{-1}$   of   LHC   data   at  a   centre-of-mass   energy   of
7~TeV~\cite{CMS:MSSM.NH},  where  $H_i$   (with  $i  =1,2,3$)  denotes
collectively  the  three  neutral  Higgs  bosons  $H_{1,2,3}$  in  the
CP-violating MSSM.   In the same section, we  also present theoretical
predictions for  $H_i \to Z \gamma$,  as these decay  modes can become
detectable  as  the integrated  LHC  luminosity increases.   Section~4
outlines how  {\tt CPsuperH2.3} incorporates calculations  of the EDMs
of Mercury and $^{225}$Ra  that include estimates of the contributions
due  to  Schiff  moments~\cite{Ellis:2011hp}.  Each  section  includes
figures  that  illustrate some  typical  results  obtained using  {\tt
  CPsuperH2.3}.   As explained  in Section~5,  an SLHA2  interface was
created  to  facilitate the  comparison  and  linkage  of output  data
between {\tt  CPsuperH2.3} and other public codes,  in accordance with
the  SUSY   Les  Houches  Accord~\cite{Skands:2003cj,Allanach:2008qq}.
Finally, Section 6 summarizes our updates to the code.

\section{Stau Contributions to Higgs masses and Mixing}
The scalar tau contributions to the masses and mixing of
the neutral Higgs  bosons have been included as in
Ref.~\cite{Choi:2000wz}, similarly to the sbottom corrections
with
\begin{equation}
X_\tau=\frac{3g^{\prime\, 2}-g^2}{4}\,
\frac{m_{\widetilde L_3}^2-m_{\widetilde E_3}^2}
{m_{\widetilde \tau_2}^2-m_{\widetilde \tau_1}^2}\,.
\end{equation}
More precisely, in the $(\phi_1,\phi_2,a)^T$ basis, the scalar tau contributions to the
entries of the neutral Higgs mass squared matrix from
the one-loop effective potential are given by
\begin{eqnarray}
\left.\left(\Delta{\cal M}_H^2\right)^{\widetilde \tau}\right|_{\phi_1\phi_1} 
&=&
\frac {m_\tau^2} {8 \pi^2} \left\{ \htau \log \frac {\mstauo \mstaut} {m_\tau^4}
- \ghat \log \frac {\mstauo \mstaut} {Q_0^4}
\right. \\ && \left. \hspace*{10mm}
+ g(\mstauo,\mstaut) R_\tau^\prime \left( \htau R_\tau^\prime + X_\tau \right) + 
\log \frac {\mstaut} {\mstauo} \left[ X_\tau + \left( 2 \htau - \ghat \right) R_\tau^\prime \right]
\right\} ; \nonumber \\[5mm]
\left.\left(\Delta{\cal M}_H^2\right)^{\widetilde \tau}\right|_{\phi_1\phi_2} 
&=&
\frac {m_\tau^2} {8 \pi^2} \left\{ g(\mstauo,\mstaut) \left[ \htau R_\tau R_\tau^\prime
+ \frac {X_\tau}{2} \left( R_\tau - R_\tau^\prime \tanb \right) \right] + \frac
{\ghat} {2} \tanb \log \frac { \mstauo \mstaut} {Q_0^4}
\nonumber \right. \\ && \left. \hspace*{10mm}
+ \log \frac {\mstaut} {\mstauo} \left[ \htau R_\tau - \frac {X_\tau}{2} \tanb +
\frac {\ghat}{2} \left( R_\tau^\prime \tanb - R_\tau \right) \right] \right\};
\\[5mm]
\left.\left(\Delta{\cal M}_H^2\right)^{\widetilde \tau}\right|_{\phi_2\phi_2} 
&=&
\frac {m_\tau^2} {8 \pi^2} \left[ g(\mstauo,\mstaut) R_\tau \left( \htau R_\tau -
\tanb X_\tau \right) + \ghat \tanb R_\tau \log \frac {\mstaut} {\mstauo} \right];
\\[5mm]
\left.\left(\Delta{\cal M}_H^2\right)^{\widetilde \tau}\right|_{a\phi_1} 
&=& \frac {1} {16 \pi^2}\,
\frac {m_\tau^2 \Delta_{\tilde \tau}} {\cos \beta} \left[ -g(\mstauo,\mstaut)
\left( X_\tau + 2 \htau R_\tau^\prime \right) + \left( \ghat - 2 \htau \right) 
\log \frac {\mstaut} {\mstauo} \right]; \\[5mm]
\left.\left(\Delta{\cal M}_H^2\right)^{\widetilde \tau}\right|_{a\phi_2} 
&=& \frac {1} {16 \pi^2}\,
\frac { m_\tau^2 \Delta_{\tilde \tau} } {\cos \beta} \left[ g(\mstauo, \mstaut)
\left( X_\tau \tanb - 2 \htau R_\tau \right) - \ghat \tanb 
\log \frac {\mstaut} {\mstauo} \right] ;
\end{eqnarray}
where $\ghat \equiv (g^2 + g'^2)/4$ and the CP-violating quantity is
\footnote{To match the {\tt CPsuperH} convention, the sign of $A_\tau$ is flipped
compared to Ref.~\cite{Choi:2000wz}.}
\begin{equation}
\Delta_{\tilde \tau} \equiv -\,\frac { \imag(A_\tau \mu) } {\mstaut - \mstauo} .
\end{equation}
The $\tan\beta$-enhanced terms containing $\mu$ and $A_\tau$ are included in
$R_\tau$ and $R_\tau^\prime$, given by
\begin{equation}
R_\tau  = \frac { \left| \mu \right|^2 \tanb - \real(A_\tau \mu )} { \mstaut - \mstauo} ; \ \ \
R_\tau^\prime = \frac { \left| A_\tau \right|^2 - \real(A_\tau \mu) \tanb} { \mstaut - \mstauo} ,
\end{equation}
and the loop function 
\begin{equation}
g(m_1^2, m_2^2) \equiv 2 - \frac {m_1^2 + m_2^2} {m_1^2 - m_2^2} \log \frac
{m_1^2} {m_2^2} .
\end{equation}
Since the stau contributions to the mass splitting $M^2_{H^\pm}-M_A^2$ are small, we have neglected them.
For the same reason,  we set
\begin{equation}
\left.\left(\Delta{\cal M}_H^2\right)^{\widetilde \tau}\right|_{aa}  =0.
\end{equation}
Note that in the limit of small mass splitting one has
\begin{equation}
\frac{g(m_1^2, m_2^2)}{(m_2^2-m_1^2)^2} \ \longrightarrow \
-\,\frac{1}{6 M_S^2}
\end{equation}
with $M_S^2=(m_1^2+m_2^2)/2$.
So, in this limit, the leading contributions proportional to $\tan^4\beta$ are
given by
\begin{eqnarray}
\left.\left(\Delta{\cal M}_H^2\right)^{\widetilde \tau}\right|_{\phi_1\phi_1} 
&\approx &  -\, \frac{m_\tau^2}{48\pi^2}\htau\, \frac{\left[\real(A_\tau \mu)\right]^2}{M_S^4}\,
\tan^2\beta ;
\nonumber \\
\left.\left(\Delta{\cal M}_H^2\right)^{\widetilde \tau}\right|_{\phi_1\phi_2} 
&\approx &  +\, \frac{m_\tau^2}{48\pi^2}\htau\, \frac{|\mu|^2\,\real(A_\tau \mu)}{M_S^4}\, \tan^2\beta ;
\nonumber \\
\left.\left(\Delta{\cal M}_H^2\right)^{\widetilde \tau}\right|_{\phi_2\phi_2} 
&\approx &  -\, \frac{m_\tau^2}{48\pi^2}\htau\, \frac{|\mu|^4}{M_S^4}\, \tan^2\beta ;
\end{eqnarray}
which are the same as those presented in 
\cite{Carena:1995bx,Carena:1995wu,Carena:2011aa}.
Note that here we take into account the
${\widetilde \tau}_{1,2}$ mass splitting.

\begin{figure}[t!]
\vspace{-1.0cm}
\begin{center}
\includegraphics[width=12.0cm]{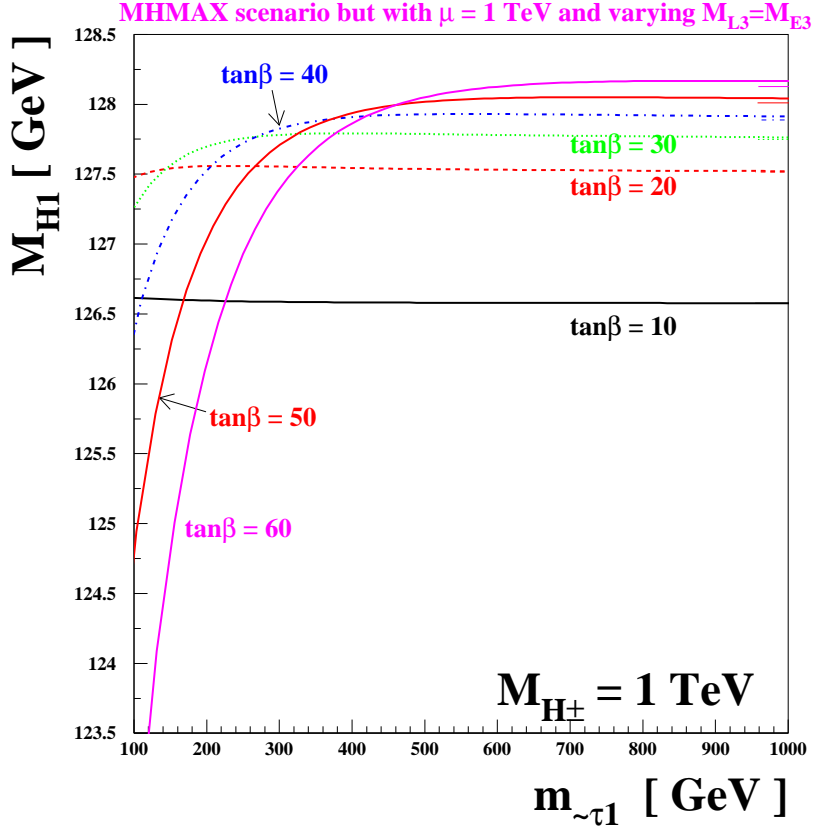} \\[-1.0cm]
\end{center}
\vspace{-0.5cm}
\caption{\it The lightest Higgs boson mass as a function of
the lighter stau mass for several values of
$\tan\beta$: 
$\tan\beta=10$ (black),
$\tan\beta=20$ (red),
$\tan\beta=30$ (green),
$\tan\beta=40$ (blue),
$\tan\beta=50$ (red), and
$\tan\beta=60$ (magenta),
as obtained varying $m_{{\widetilde L}_3}=m_{{\widetilde E}_3}$ with fixed $\mu=1$ TeV.
We use the {\tt MHMAX} scenario for the other supersymmetric model parameters:
$m_{\tilde{Q}_3}= m_{\tilde{U}_3}= m_{\tilde{D}_3}= M_{\rm SUSY} = 1~{\rm TeV}$,
$M_1=100~{\rm GeV}\,, M_2=200~{\rm GeV}\,, M_3=800~{\rm GeV}$, and
$A_t=\sqrt{6}\,M_{\rm SUSY}+\mu/\tan\beta$ with $A_b=A_\tau=A_t$.
We have fixed $M_{H^\pm}=1$ TeV.
The thin short line
around $m_{{\widetilde \tau}_1}= 1~{\rm TeV}$
for each $\tan\beta$ value shows
the lightest Higgs boson mass before the inclusion of the stau effects.
}
\label{fig:mh1.mstu1}
\end{figure}
In Fig.~\ref{fig:mh1.mstu1}, we demonstrate the impact of the 
corrections from the light scalar taus for large $\mu = 1$ TeV for several values of
$\tan\beta$. 
When $m_{{\widetilde \tau}_{1,2}}\sim 1$ TeV,
the stau corrections are negligible and the lightest Higgs boson mass increases
by the small amount of
$\sim 0.05$ GeV even when $\tan\beta = 60$. 
However, these corrections are larger for lighter staus, and the
lightest Higgs boson mass may decrease by as much as $\sim 4$~GeV, if
$\tan\beta =60$ and $m_{\widetilde\tau_1} \sim 100$ GeV.
For comparison, as shown in Table~\ref{tab:raux_2},
the masses and mixing matrix of the neutral Higgs bosons without including 
the stau effects
are stored in the array {\tt RAUX\_H}:
\begin{itemize}
\item[-]
{\tt RAUX\_H(511 $-$ 513)} : Three Higgs masses without including
the stau effects;
\item[-]
{\tt RAUX\_H(520 $-$ 528)} : The nine elements of the mixing matrix $O_{\alpha i}$
without including the stau effects.
\end{itemize}

\section{Collider Limits}
The limits on Higgs boson production at LEP and the LHC experiments have been implemented
as follows.

\subsection{LEP Limits}
We have implemented the LEP limits from the processes
$e^+e^-\to H_i Z$ and $e^+e^-\to H_i H_j$ using
Table 14\,(a) and Table 17\,(a) of Ref.~\cite{Schael:2006cr},
respectively, assuming that the Higgs decay patterns are
not drastically different from those in the Standard Model and
in the {\tt MHMAX} scenario with $\tan\beta=10$.
We have required that each of the
Higgs couplings to the gauge boson(s) normalized to the Standard Model values, 
$g_{H_iVV}^2$ and $g_{H_iH_jZ}^2$, should be smaller than the corresponding values
in the Tables. The result is saved in
\begin{itemize}
\item[-]
{\tt RAUX\_H(430)}$=$ILEP : 0 (Excluded) or 1 (Allowed).
\end{itemize}
For the more general cases, we refer to more refined tools such as
{\tt HiggsBounds}~\cite{Bechtle:2008jh}.

\subsection{LHC Limits}
We have also incorporated the recent CMS limit from
the search for the MSSM neutral Higgs bosons decaying into tau pairs
based on 4.6 fb$^{-1}$~\cite{CMS:MSSM.NH}, and the result is saved in
\begin{itemize}
\item[-]
{\tt RAUX\_H(440)}$={\rm ILHC7}^{H\to\tau\tau}_{4.6}$ : 0 (Excluded) or 1 (Allowed) .
\end{itemize}
%
In our implementation of the CMS limit, 
assuming the same K factors as in the SM,
we approximate the production cross sections of the 
neutral Higgs bosons at the LHC as follows:
\begin{equation}
\sigma_{\cal P}^{\rm MSSM}(pp\to H_i\,X) \ \simeq \
\left(\frac{\Gamma_{\cal P}^{\rm MSSM}}{\Gamma_{\cal P}^{\rm SM}}\right)^{LO}_i
\left.
\sigma_{\cal P}^{\rm SM}(pp\to H_{\rm SM}\,X) \right|_{M_{H_{\rm SM}}=M_{H_i}} ,
\end{equation}
where ${\cal P}=ggH_i, bbH_i, VVH_i$ specifies each production process.  
The SM production cross sections 
are calculated by using 
{\tt HIGLU} \cite{Spira:1995mt}, 
{\tt BBH@NNLO} \cite{Harlander:2003ai} and
{\tt HAWK} \cite{hawk}
\footnote{We thank Junya Nakamura for 
providing the SM cross sections.}.
In leading order,
the process-dependent ratios are given by:
\begin{itemize}
\item[$-$] \underline{Gluon fusion}
\begin{equation}
\left(\frac{\Gamma_{ggH_i}^{\rm MSSM}}{\Gamma_{ggH_i}^{\rm SM}}\right)^{LO}_i \ = \
\frac{|S_i^g(M_{H_i})|^2+|P_i^g(M_{H_i})|^2}{|S_{\rm SM}^g(M_{H_i})|^2} 
\ \equiv \ R_{H_igg} ,
\label{eq:fac_gg}
\end{equation}
where $S_{\rm SM}^g(M_{H_i})=
\sum_{f=b,t}\,F_{sf}(\tau_{if})$. These factors are saved in the array {\tt CAUX\_H(221-223)}
as shown in Table~\ref{tab:caux_2}.
\item[$-$] \underline{$b$-quark fusion}
\begin{equation}
\left(\frac{\Gamma_{bbH_i}^{\rm MSSM}}{\Gamma_{bbH_i}^{\rm SM}}\right)^{LO}_i \ = \
\left|g^S_{H_i\bar{b}b}\right|^2+
\frac{\left|g^P_{H_i\bar{b}b}\right|^2}{(1-4m_b^2/M_{H_i}^2)}
\ \equiv \ R_{H_ibb} .
\label{eq:fac_bfusion}
\end{equation}
where the extra factor in the second term accounts for the phase-space difference 
between scalar and pseudo-scalar decays.
{\color{blue}In the program, we 
drop the extra factor
since the ratio is involved in production process.}
\item[$-$] \underline{Vector-boson fusion}
\begin{equation}
\left(\frac{\Gamma_{VVH_i}^{\rm MSSM}}{\Gamma_{VVH_i}^{\rm SM}}\right)^{LO}_i \ = \
g_{H_iVV}^2 .
\label{eq:fac_VBF}
\end{equation}
\end{itemize}
\begin{figure}[t!]
\vspace{-1.0cm}
\begin{center}
\includegraphics[width=12.0cm]{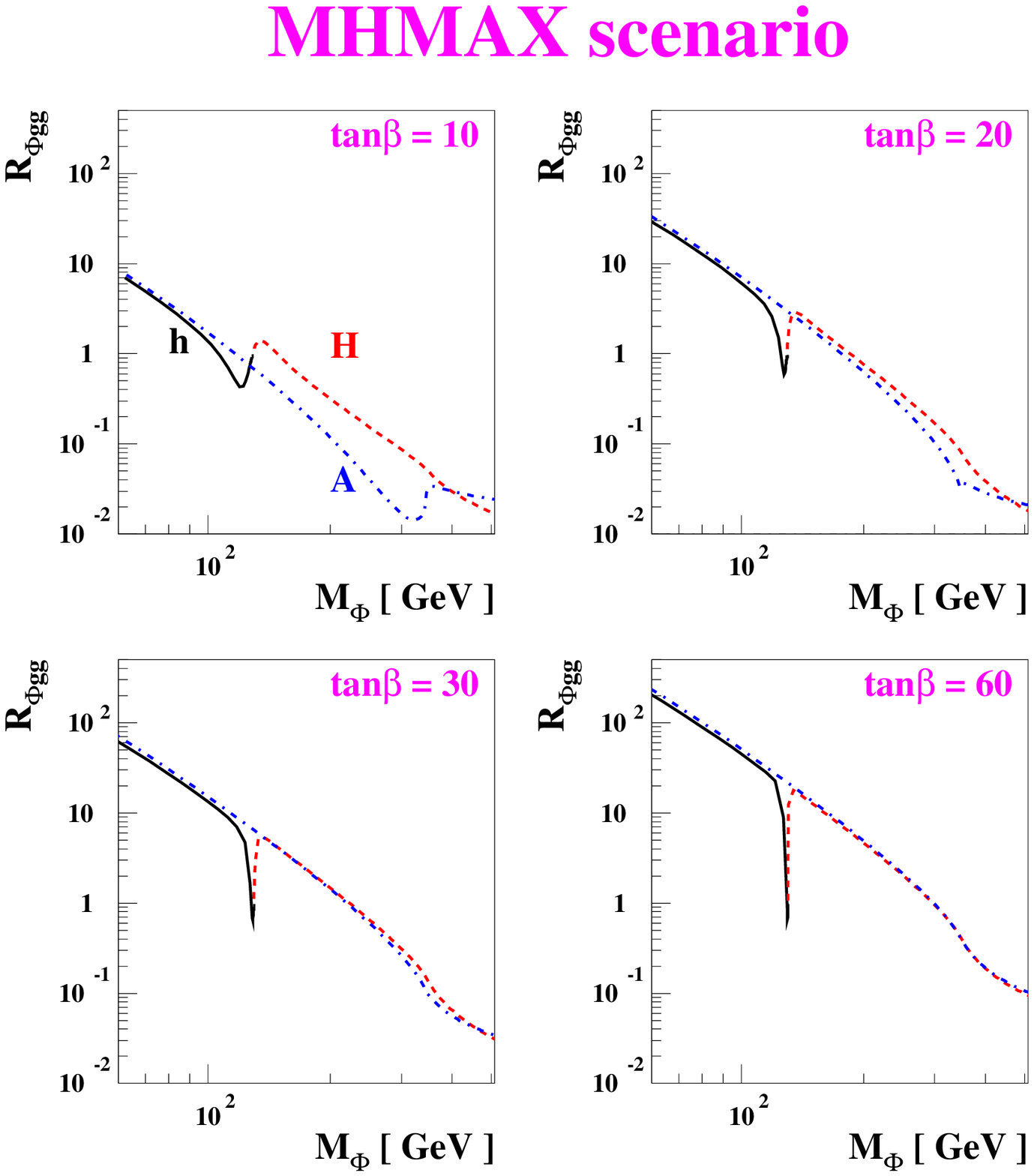} \\[-1.0cm]
\end{center}
\vspace{-0.5cm}
\caption{\it The couplings of the MSSM Higgs bosons $\Phi$ to two gluons 
normalized to the Standard Model values, $R_{H_i gg}$ (\protect\ref{eq:fac_gg}), 
as functions of their masses for several values of $\tan\beta$.
In the CP-conserving limit,
$\Phi \equiv h,H,A$ with $h (H)$ and $A$ being the lighter (heavier) CP-even and CP-odd Higgs bosons,
respectively. We assume the {\tt MHMAX} scenario:
$m_{\tilde{Q}_3}= m_{\tilde{U}_3}= m_{\tilde{D}_3}= 
m_{\tilde{L}_3}= m_{\tilde{E}_3}= M_{\rm SUSY} = 1~{\rm TeV}$,
$\mu=200~{\rm GeV}\,,
M_1=100~{\rm GeV}\,, M_2=200~{\rm GeV}\,, M_3=800~{\rm GeV}$, and
$A_t=\sqrt{6}\,M_{\rm SUSY}+\mu/\tan\beta$ with $A_b=A_\tau=A_t$.}
\label{fig:ghgg}
\end{figure}
\begin{figure}[t!]
\vspace{-1.0cm}
\begin{center}
\includegraphics[width=12.0cm]{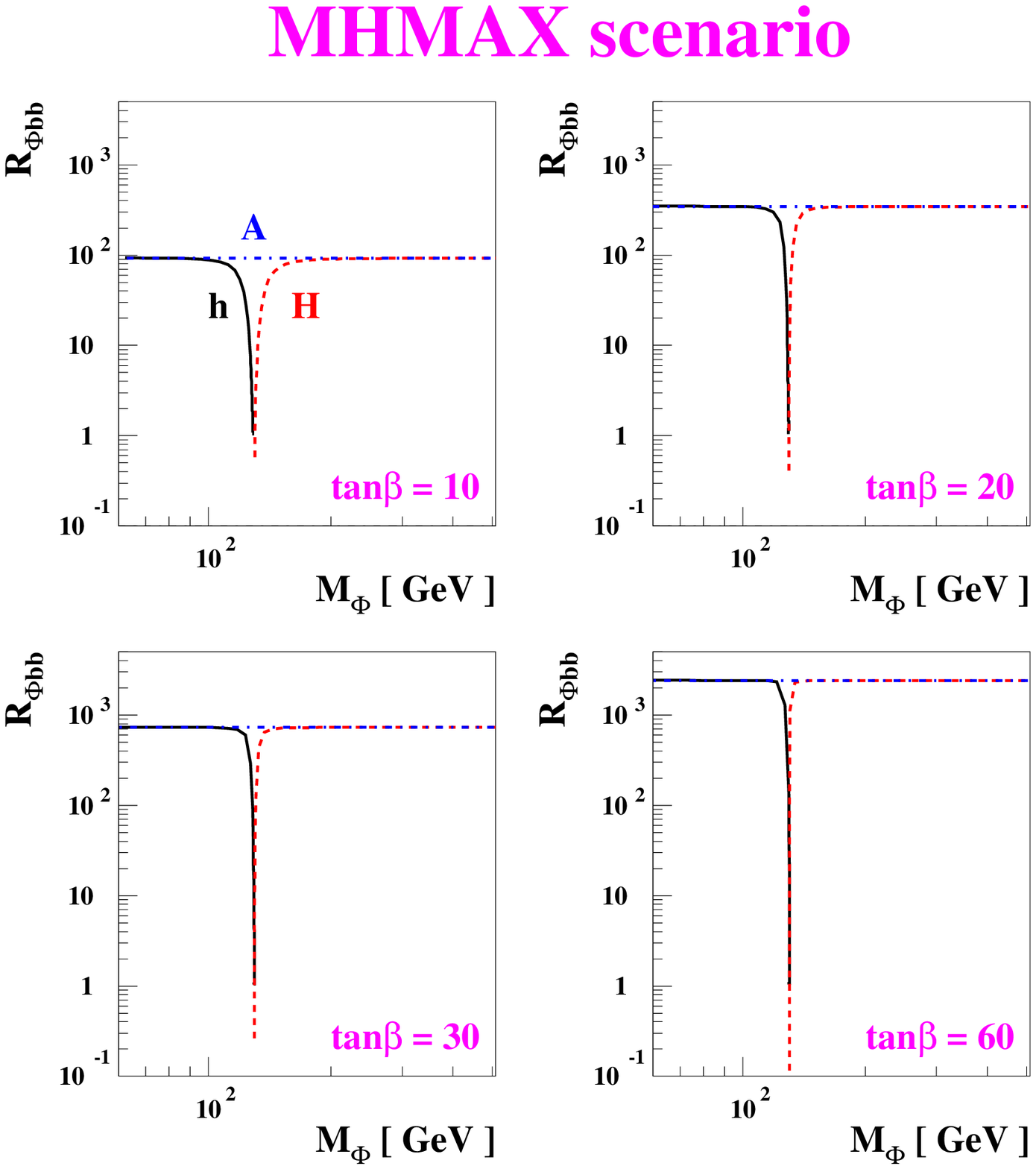} \\[-1.0cm]
\end{center}
\vspace{-0.5cm}
\caption{\it The same as in Fig.~\ref{fig:ghgg} but for the 
couplings of the Higgs bosons to $b$ quarks (\protect\ref{eq:fac_bfusion}).}
\label{fig:ghbb}
\end{figure}
\begin{figure}[t!]
\vspace{-1.0cm}
\begin{center}
\includegraphics[width=12.0cm]{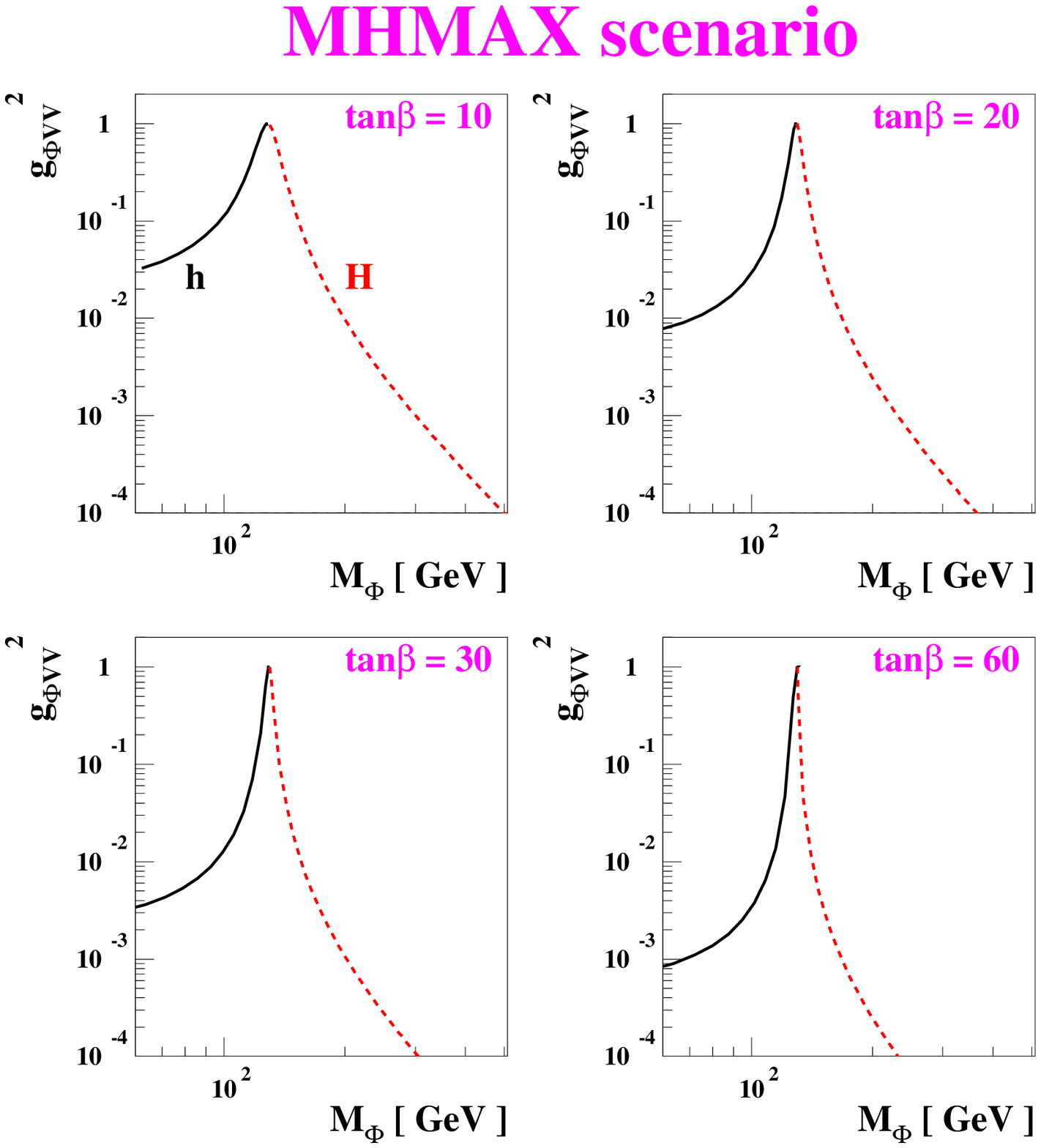} \\[-1.0cm]
\end{center}
\vspace{-0.5cm}
\caption{\it The same as in Fig.~\ref{fig:ghgg} but for the 
couplings of the Higgs bosons to two vector bosons (\protect\ref{eq:fac_VBF}).}
\label{fig:ghvv}
\end{figure}
\begin{figure}[t!]
\vspace{-1.0cm}
\begin{center}
\includegraphics[width=12.0cm]{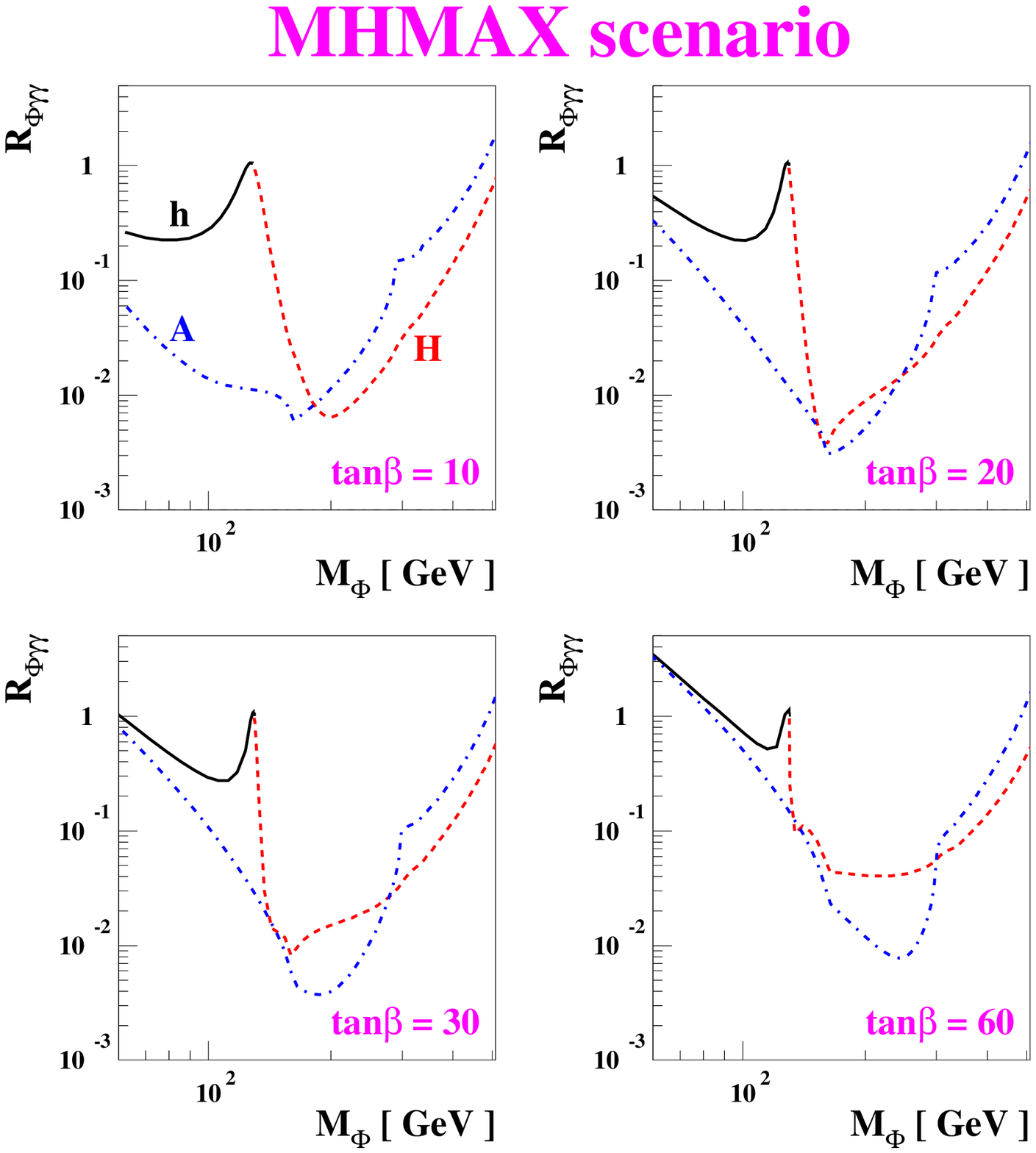} \\[-1.0cm]
\end{center}
\vspace{-0.5cm}
\caption{\it The same as in Fig.~\ref{fig:ghgg} but for the
couplings of the Higgs bosons to two photons.}
\label{fig:ghpp}
\end{figure}
\begin{figure}[t!]
\vspace{-1.0cm}
\begin{center}
\includegraphics[width=12.0cm]{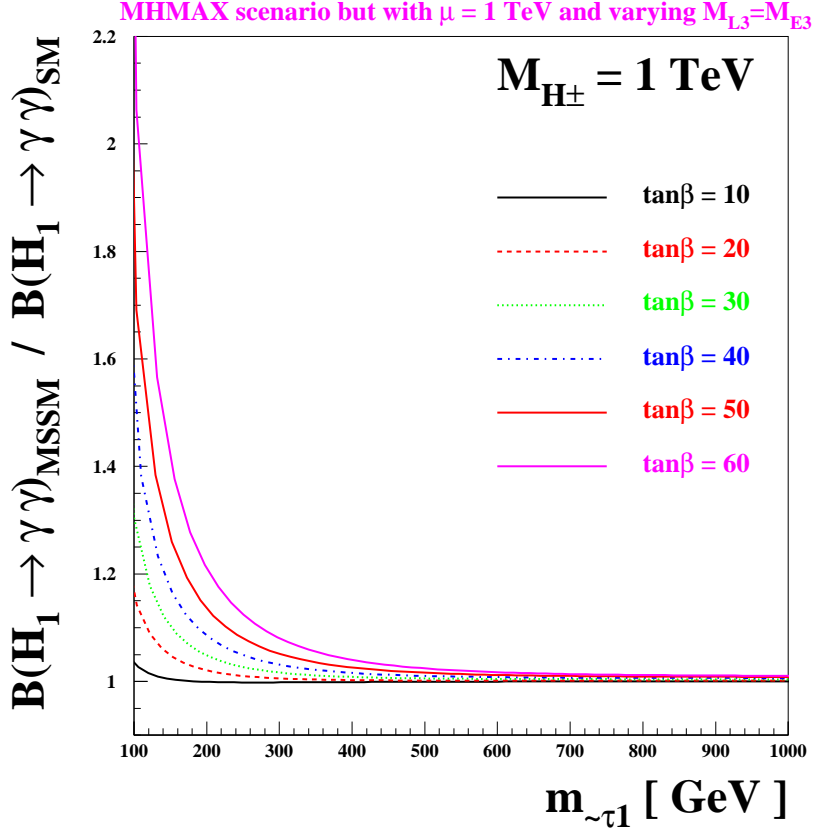} \\[-1.0cm]
\end{center}
\vspace{-0.5cm}
\caption{\it The ratio of the MSSM-to-SM
branching ratios of the lightest Higgs boson
$H_1$ decaying into photons as a function of the lighter stau mass.
The scenario and
lines are the same as in Fig.~\ref{fig:mh1.mstu1}.
}
\label{fig:rpp.mstu1}
\end{figure}
\begin{figure}[t!]
\vspace{-1.0cm}
\begin{center}
\includegraphics[width=12.0cm]{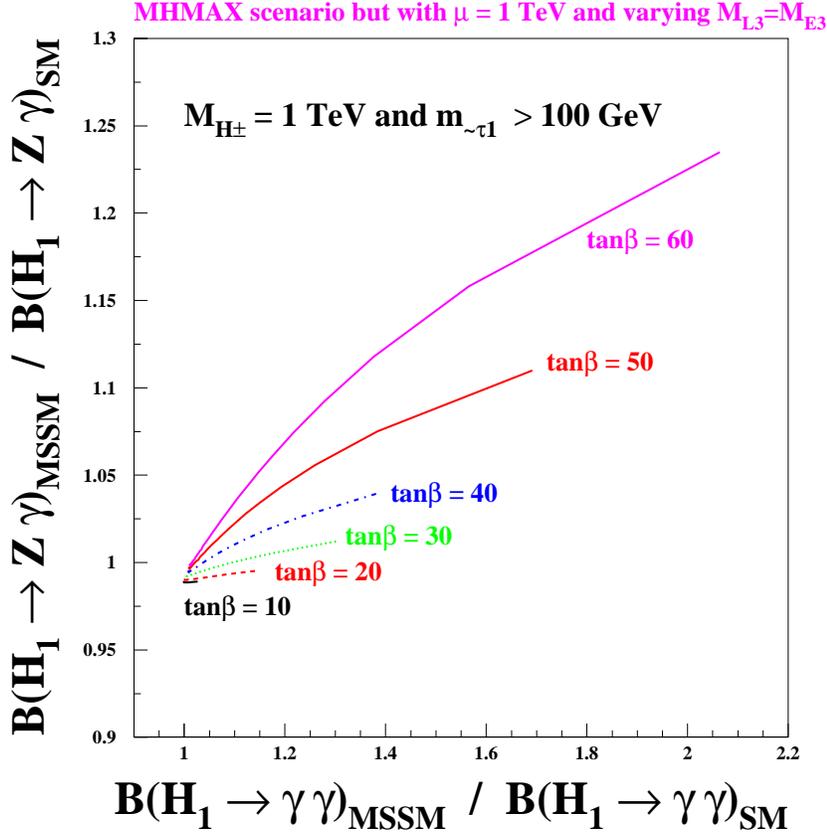} \\[-1.0cm]
\end{center}
\vspace{-0.5cm}
\caption{\it  Correlations of  the MSSM-to-SM  branching ratio  of the
lightest  Higgs  boson  $H_1$  decaying  into  two  photons  with  the
respective one for the decay $H_1$  into a $Z$ boson and a photon, for
discrete choices of $\tan\beta$.  The  scenario and lines are the same
as in Fig.~\ref{fig:mh1.mstu1}.  }
\label{fig:rpp.rzp}
\end{figure}
\begin{figure}[t!]
\vspace{-1.0cm}
\begin{center}
\includegraphics[width=12.0cm]{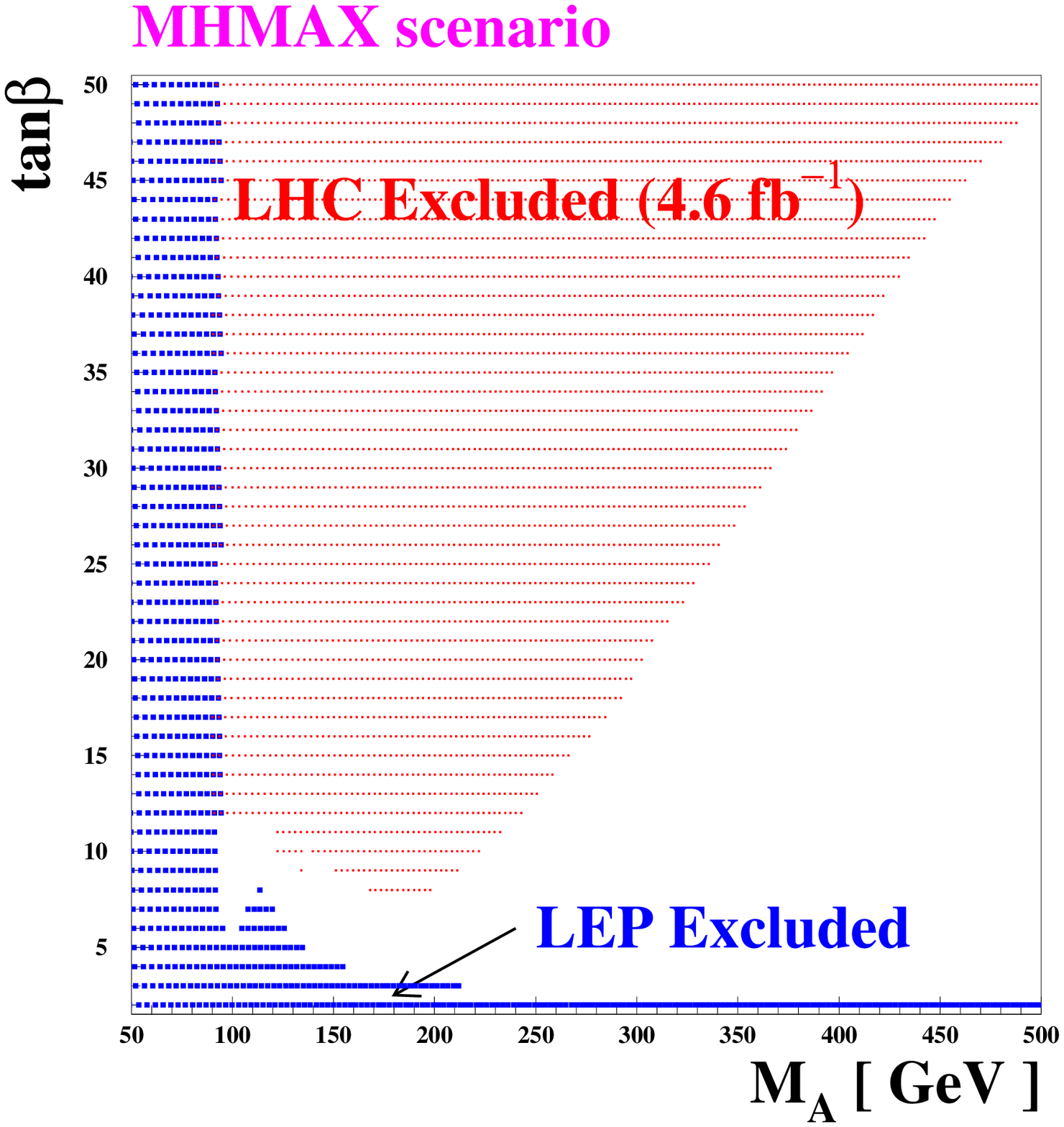} \\[-1.0cm]
\end{center}
\vspace{-0.5cm}
\caption{\it The LEP and LHC exclusion regions in the {\tt MHMAX} scenario.}
\label{fig:coll} 
\end{figure}
In Figs.~\ref{fig:ghgg},\ref{fig:ghbb},\ref{fig:ghvv}, we show the three
process-dependent ratios as functions of the corresponding Higgs masses,
assuming the {\tt MHMAX} scenario. 
For completeness, also shown in Fig.~\ref{fig:ghpp} are
the ratios of the Higgs-boson couplings to two photons
defined by
\begin{eqnarray}
R_{H_i\gamma\gamma} &\equiv & \frac{|S_i^\gamma(M_{H_i})|^2
+|P_i^\gamma(M_{H_i})|^2}{|S_{\rm SM}^\gamma(M_{H_i})|^2} ,
\end{eqnarray}
where $S_{\rm SM}^\gamma(M_{H_i}) \equiv 2
\sum_{f=b,t,c,\tau}N_C\,Q_f^2\,F_{sf}(\tau_{if})
-F_{sf}(\tau_{iW})$,
which are saved in the array {\tt CAUX\_H(231-233)}
as shown in Table~\ref{tab:caux_2}.
%

We observe that the branching  ratio $B(H_1 \to \gamma \gamma)$ may be
enhanced if  the lighter stau has  a low mass,  particularly for large
$\tan \beta$  as shown in  Fig.~\ref{fig:rpp.mstu1}.  This enhancement
is consistent with the  signals observed \cite{ATLASCMS}, which may be
larger    than    in   the    Standard    Model.    Furthermore,    in
Fig.~\ref{fig:rpp.rzp} we  show the  correlation of $B(H_1  \to \gamma
\gamma)$ with  $B(H_1 \to Z\gamma)$  in the MSSM, after  the branching
ratios are normalized to  their SM predictions.  Calculational details
of $B(H_i \to Z \gamma)$ are described in Appendix~A.

Once the  production cross sections  of the neutral MSSM  Higgs bosons
have  been  calculated,  we   may  require   the  sum~\footnote{In  the
  CP-conserving  limit,  the CP-odd  state  $A$  is  $H_i$ with  $O_{a
    i}^2=1$.   In  the CP-violating  case,  we  identify the  ``CP-odd
  state" as the one which has the the largest CP-odd component squared
  $O_{a i}^2$.}
\begin{equation}
\sigma(pp\to H_i X)\cdot B(H_i \to \tau\tau) +
\left.
\sum_{\phi\neq H_i}\,\right|_{|M_\phi-M_{H_i}|\leq\delta M}
\,\sigma(pp\to \phi X)\cdot B(\phi\to \tau\tau)
\end{equation}
to be smaller than the observed limit for a given value $M_{H_i}$ of a
specified  neutral Higgs boson~$H_i$.   Here we  are adding  all Higgs
production cross sections  of $\phi \neq H_i$ to the  one of the $H_i$
boson when their mass difference  is smaller than $\delta M$.  We take
$\delta M=0.21*130/2 \sim 13$~GeV,  corresponding to the tau-pair mass
resolution  of   $\sim  21$   \%  at  a   Higgs  boson  mass   of  130
GeV~\cite{CMS:MSSM.NH}.  Such  a simple treatment  is fairly accurate,
as  long as  $\delta M  \gg \Gamma_{H_i}$,  namely in  the  absence of
strongly overlapping Higgs resonances~\cite{Ellis:2004fs}. Under these
considerations,  Fig.~\ref{fig:coll} shows  the LEP-  and LHC-excluded
regions in  the $M_A$-$\tan\beta$ plane for the  {\tt MHMAX} scenario.
We  find our  results reasonably  consistent with  those  presented in
Fig.~4 of~\cite{CMS:MSSM.NH} when $M_A\geq 90$ GeV.

\begin{figure}[t!]
\vspace{-1.0cm}
\begin{center}
\includegraphics[width=12.0cm]{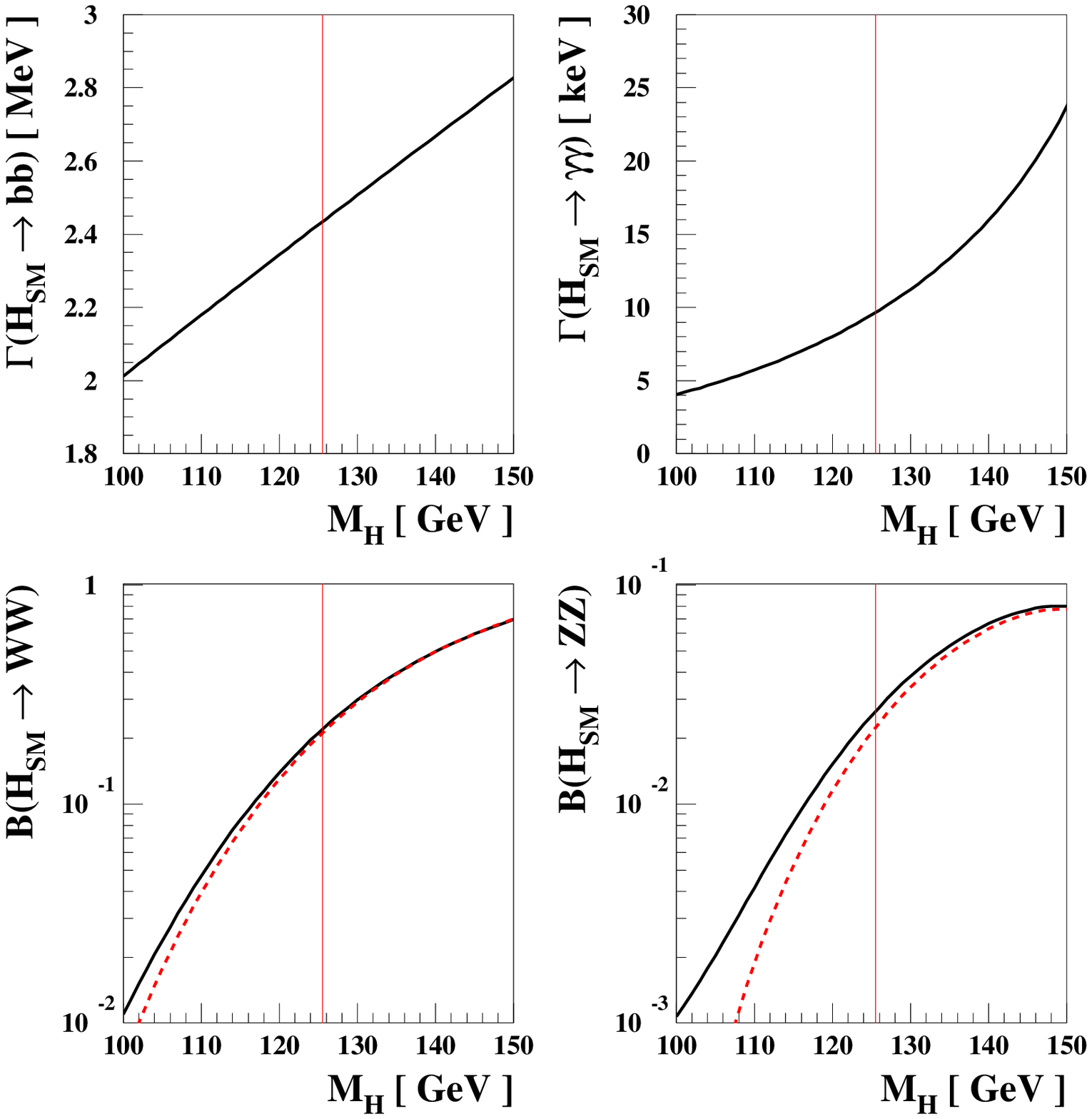} \\[-1.0cm]
\end{center}
\vspace{-0.5cm}
\caption{\it  The upper frames are for the decay widths 
of the SM Higgs boson into $b$ quarks (upper left)
and photons (upper right) as functions of its mass. 
The lower frames are for the branching ratios 
into $W$ (lower left) and $Z$ (lower right) bosons.
In the lower frames, the solid (dotted) lines are obtained using
4\,(3)-body phase space below the mass thresholds.
The vertical lines correspond to $M_{H_{\rm SM}}=125.5$~GeV.
}
\label{fig:smbrs}
\end{figure}

Finally, 
for the value given by $M_{H_{\rm SM}}=${\tt SMPARA\_H(20)},
the decay widths and branching ratios of the SM Higgs have
been calculated, and the results are stored in {\tt RAUX\_H(600-700)},
see Table \ref{tab:raux_2}.  For comparison with other works, we
display in Fig.~\ref{fig:smbrs} the decay widths 
into $b$ quarks and photons  (upper)
and the branching ratios
into $W$ and $Z$ bosons (lower)
as functions of the SM Higgs-boson mass.  
We note that decay widths into off-shell
vector bosons using the four-body phase-space have been implemented in
{\tt CPsuperH2.3}.

\section{Mercury and Radium Electric Dipole Moments}
\begin{figure}[t!]
\vspace{-1.0cm}
\begin{center}
\includegraphics[width=12.0cm]{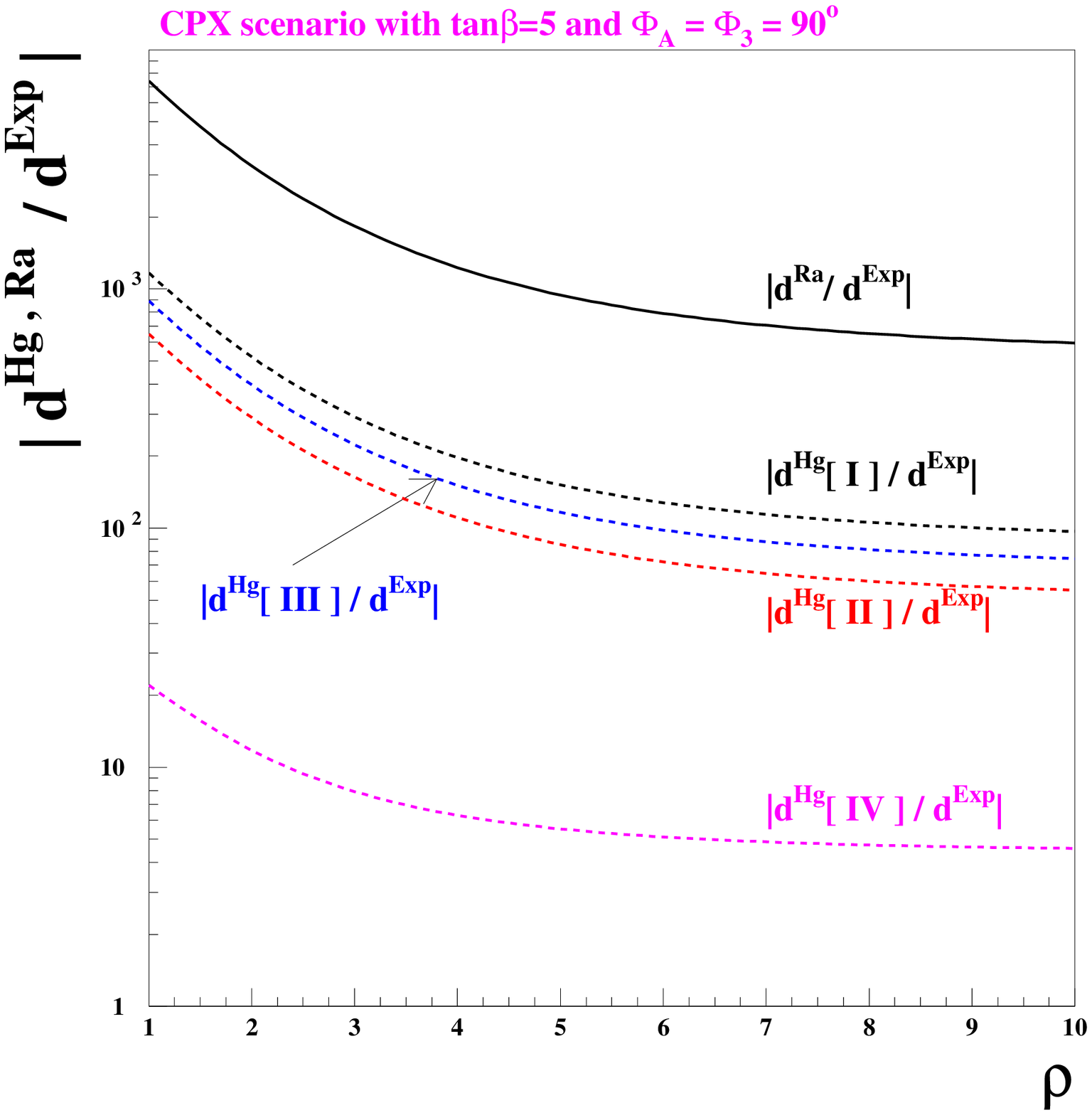} \\[-1.0cm]
\end{center}
\vspace{-0.5cm}
\caption{\it The four Mercury EDMs and the Radium EDM
as functions of the parameters $\rho$ that parameterizes the hierarchy between the first two 
and third generations as $m_{\tilde{X}_{1,2}}=\rho\, m_{\tilde{X}_3} $
with $X=Q,U,D,L,E$.
The CPX scenario has been taken with $\tan\beta=5$, $\Phi_{A_{t,b}}=\Phi_3=90^\circ$,
and $M_{\rm SUSY}=0.5~~{\rm TeV}$:
$m_{\tilde{Q}_3} = m_{\tilde{U}_3} = m_{\tilde{D}_3} =
m_{\tilde{L}_3} = m_{\tilde{E}_3} = M_{\rm SUSY}$;
$|\mu|=4\,M_{\rm SUSY}\,,
|A_{t,b,\tau}|=2\,M_{\rm SUSY} \,, 
|M_3|=1$ TeV;
$M_2=2M_1=100~~{\rm GeV}\,,
M_{H^\pm}=300~~{\rm GeV}$.
}
\label{fig:rahg}
\end{figure}

{\tt CPsuperH2.0} included two-loop Higgs-mediated contributions to the
Thallium and electron EDMs~\cite{Lee:2007gn}, and {\tt CPsuperH2.3}
extends these results to include calculations of the Mercury and $^{225}$Ra
EDMs that incorporate Schiff-moment contributions.
In the case of the Mercury EDM, these are parameterized as
follows~\cite{Ellis:2008zy}: 
\begin{eqnarray}
d^{\rm \,I\,,II\,,III\,,IV}_{\rm Hg} \!&=&\!
d^{\rm \,I\,,II\,,III\,,IV}_{\rm Hg}[S]
+10^{-2} d_e^E 
+(3.5\times 10^{-3} {\rm GeV})\,e\,C_S
\nonumber \\ \!&&\!
+\ (4\times 10^{-4}~{\rm GeV})\,e\,\left[C_P+
\left(\frac{Z-N}{A}\right)_{\rm Hg}\,C^\prime_P\right] , \nonumber
\end{eqnarray}
where $d^{\rm \,I\,,II\,,III\,,IV}_{\rm Hg}[S]$
denote the following four different Schiff-moment induced Mercury EDM 
calculations:
\begin{eqnarray}
d^{\rm \,I}_{\rm Hg}[S]\ & \simeq\ &
\hspace{4.7cm}
1.8 \times 10^{-3}\, e\,\bar{g}^{(1)}_{\pi NN}\,/{\rm GeV}\,,
\nonumber \\
d^{\rm \,II}_{\rm Hg}[S]\ & \simeq\ &
7.6 \times 10^{-6}\, e\,\bar{g}^{(0)}_{\pi NN}\,/{\rm GeV}+
1.0 \times 10^{-3}\, e\,\bar{g}^{(1)}_{\pi NN}\,/{\rm GeV}\,,
\nonumber \\
d^{\rm \,III}_{\rm Hg}[S]\ & \simeq\ &
1.3 \times 10^{-4}\, e\,\bar{g}^{(0)}_{\pi NN}\,/{\rm GeV}+
1.4 \times 10^{-3}\, e\,\bar{g}^{(1)}_{\pi NN}\,/{\rm GeV}\,,
\nonumber \\
d^{\rm \,IV}_{\rm Hg}[S]\ & \simeq\ &
3.1 \times 10^{-4}\, e\,\bar{g}^{(0)}_{\pi NN}\,/{\rm GeV}+
9.5 \times 10^{-5}\, e\,\bar{g}^{(1)}_{\pi NN}\,/{\rm GeV} , \nonumber
\end{eqnarray}
where the $\bar{g}^{(0),(1)}_{\pi NN}$ are the CP-odd $\pi NN$ couplings, 
and we refer to Ref.~\cite{Ellis:2011hp} for further details. We note that
$d^{\rm \,I}_{\rm Hg}$ is basically the same as that calculated in
{\tt CPsuperH2.0}.
In the case of the Radium $^{225}$Ra, we estimate the EDM through~\cite{Ellis:2011hp}
\begin{eqnarray}
d_{\rm Ra} \simeq
d_{\rm Ra}[S] \simeq
-8.7 \times 10^{-2}\, e\,\bar{g}^{(0)}_{\pi NN}\,/{\rm GeV}
+3.5 \times 10^{-1}\, e\,\bar{g}^{(1)}_{\pi NN}\,/{\rm GeV}\,.
\end{eqnarray}
Fig.~\ref{fig:rahg} shows the four estimates of the Mercury EDM and the Radium EDM
as functions of the parameter $\rho$ that parameterizes the hierarchy between the first two
and third generations in the {\tt CPX} scenario. The new EDMs are stored in
the array {\tt RAUX\_H}:
\begin{itemize}
\item[-]
{\tt RAUX\_H(411)} $=d^{\rm \,I}_{\rm Hg}$,
{\tt RAUX\_H(412)} $=d^{\rm \,II}_{\rm Hg}$ ,
{\tt RAUX\_H(413)} $=d^{\rm \,III}_{\rm Hg}$,
{\tt RAUX\_H(414)} $=d^{\rm \,IV}_{\rm Hg}$ 
\item[-]
{\tt RAUX\_H(420)} $=d_{\rm Ra}$ 
\end{itemize}
as shown in Table~\ref{tab:raux_1}.

\section{{\tt SLHA2} interface
\protect\footnote{We thank Alexander Pukhov for 
helpful discussions regarding the interface.}
}
{\tt CPsuperH2.3} also provides
an output in accordance with the SUSY Les Houches Accords 1 
(SLHA1)~\cite{Skands:2003cj}
and 2 (SLHA2)~\cite{Allanach:2008qq}. 
By taking 
\footnote{Note that the default setting
is {\tt IFLAG\_H(30)}=0, which is not recommended for generating files
when a scan of the parameter space is performed.}
\begin{itemize}
\item {\tt IFLAG\_H(30)}=1 
\end{itemize}
in the {\tt run} file, the output file 
{\tt cpsuper2.3\_slha2.out} is generated.
The output file of the current version
includes the following blocks:
\begin{itemize}
\item {\tt MODSEL}:
In the block {\tt MODSEL}, CP violation with completely
general CP phases has been selected. 
\item {\tt SMINPUTS} and {\tt VCKMIN}: The quantities
$\alpha_{\rm em}^{-1}(M_Z)^{\overline{\rm MS}}$, $G_F$,
$\alpha_s(M_Z)^{\overline{\rm MS}}$, $M_Z$,
$m_b(m_b)^{\overline{\rm MS}}$, 
$m_c(m_c)^{\overline{\rm MS}}$, 
pole masses of top-quark, electron and muon, and
$m_{d,u,s}(2~{\rm GeV})^{\overline{\rm MS}}$ are given in the {\tt SMINPUTS}
block. In the {\tt VCKMIN} block, the four parameters for the CKM mixing matrix
$\lambda$, $A$, $\bar\rho$, and $\bar\eta$ are given.
\item {\tt EXTPAR} and {\tt IMEXTPAR}: The blocks  for non-minimal parameters have been
filled according to Section 2 of the SLHA2 writeup~\cite{Allanach:2008qq} 
\footnote{For the Higgs parameters, only
the relevant quantities $\mu(M_{\rm input})$,
$\tan\beta(M_{\rm input})$, and the charged Higgs pole mass have been included.}.
For the input scale $M_{\rm input}$,
we are taking $Q_{tb}$, the scale of  the heaviest third-generation
squark~\cite{Carena:2000yi}, which is stored in {\tt RAUX\_H(13)}$=Q_{tb}^2$.
The imaginary parts
of the gaugino masses, the trilinear couplings, and the $\mu$
parameters are listed in the block {\tt IMEXTPAR}.
\item {\tt MASS}: In the {\tt MASS} block, the pole masses of the neutral and charged Higgs
bosons are given, and the $W$-boson mass is calculated from the Fermi constant via
$M_W=\left(g^2/4\sqrt{2}G_F\right)^{1/2}$. 
Note that, in the present version, the treel-level masses of the
third-generation sfermions, neutralinos, and charginos
are given.
\item {\tt HCOUPLINGS},
{\tt IMHCOUPLINGS}: In these blocks, the seven
Higgs-self couplings $\lambda_{1-7}$ calculated according to
Ref.~\cite{Pilaftsis:1999qt}
are given from {\tt CAUX\_H(201$-$207)}.
\item {\tt THRESHOLD}: In this block, the measures of the threshold corrections
to the Yukawa couplings $h_{t,b}$
are given from {\tt CAUX\_H(211,212)}. The quantity $\Delta_{b}$ is also available
from {\tt CAUX\_H(10)}.
\item {\tt NMIX},
{\tt IMNMIX};
{\tt UMIX},
{\tt IMUMIX},
{\tt VMIX},
{\tt IMVMIX};
{\tt STOPMIX},
{\tt IMSTOPMIX};
{\tt SBOTMIX},
{\tt IMSBOTMIX};
{\tt STAUMIX},
{\tt IMSTAUMIX}: In these blocks, the mixing matrices of neutralinos, charginos, sfermions
are given. Precisely, 
\begin{itemize}
\item $\left({\tt NMIX} + I\,  {\tt IMNMIX}\right)_{i\alpha} = N_{i\alpha}$
with $i=1, 2, 3, 4$ and $\alpha=(\widetilde{B},\widetilde{W}^3,\widetilde{H}^0_1,\widetilde{H}^0_2)$ ,
\item $\left({\tt UMIX} + I\,  {\tt IMUMIX}\right)_{i\alpha} = 
\left(C_L\right)_{i\alpha}$
with $i=1, 2$ and $\alpha=(\widetilde{W}^-,\widetilde{H}^-)_L$ ,
\item $\left({\tt VMIX} + I\,  {\tt IMVMIX}\right)_{i\alpha} = 
\left(C_R\right)^*_{i\alpha}$
with $i=1, 2$ and $\alpha=(\widetilde{W}^-,\widetilde{H}^-)_R$ ,
\item $\left({\tt STOPMIX} + I\,  {\tt IMSTOPMIX}\right)_{i\alpha} = 
\left(U^{\tilde t}\right)^*_{\alpha i}$
with $i=1, 2$ and $\alpha=(\widetilde{t}_L,\widetilde{t}_R)$,
and similarly for sbottoms and staus.
\end{itemize}
\item {\tt CVHMIX}: The block for the neutral Higgs boson mixing has
  been filled as 
suggested in Ref.~\cite{Allanach:2008qq}:
\begin{equation}
\left({\tt CVHMIX}\right)_{i\sigma} = \left(
\begin{array}{c c}
    & 0 \\
O^T_{3\times 3} & 0 \\
   & 0 \end{array} \right)_{i\alpha} \
\left(
\begin{array}{cccc}
1 & 0 & 0 & 0 \\
0 & 1 & 0 & 0 \\
0 & 0 & -\sin\beta & \cos\beta \\
0 & 0 & \cos\beta & \sin\beta \end{array} \right)_{\alpha\sigma} ,
\end{equation}
with $i=1, 2, 3$ ,
$\alpha=(\phi_1,\phi_2,a,G^0)$
and $\sigma=(\phi_1,\phi_2,a_1,a_2)$. We note that
$a=-\sin\beta\, a_1 + \cos\beta\, a_2$ and
$G^0=\cos\beta\, a_1 + \sin\beta\, a_2$.
\item {\tt AU},
{\tt IMAU},
{\tt AD},
{\tt IMAD},
{\tt AE},
{\tt IMAE;}
{\tt YU},
{\tt IMYU},
{\tt YD},
{\tt IMYD},
{\tt YE},
{\tt IMYE}:
In these blocks, the third-generation $(3,3)$ components of the
trilinear-coupling  
and the Yukawa-coupling matrices are
given at the scales $Q_{tb}$ and $m_t^{\rm pole}$, respectively.
\item {\tt DECAY}: The total decay widths and the
non-vanishing branching ratios of the neutral and charged Higgs bosons
are stored in the block from the arrays
{\tt GAMBRN(101,1,IH)} and
{\tt GAMBRN(IM,3,IH)} with {\tt IH}$=1-3$ and 
{\tt GAMBRC(51,1)} and
{\tt GAMBRC(IM,3)}, respectively.
The decay width and branching ratios of the top quark
are also stored from the arrays {\tt RAUX\_H(50-53)}. 

\item {\tt FOBS}: In this block, we show 
the branching ratios
$B(b\to X_s\gamma)$,
$B(B_s\to \mu\mu)$,
$B(B_d\to \tau\tau)$, 
the ratio $R_{B\tau\nu}$, and
the CP asymmetry ${\cal A}_{\rm CP}(B\to X_s\gamma)$.
These quantities are also available from
{\tt RAUX\_H(130,131)} and
{\tt RAUX\_H(134,135,136)} in different normalizations.
\item {\tt FOBSBSM}: In this block, we show the SUSY contributions
to $\Delta M_{B_d,B_s}$ from
{\tt RAUX\_H(132,133)}. For the total $\Delta M_{B_d,B_s}$,
one may add the quantities
$\bra{\bar{B}^0_d}\, H_{\rm eff}^{\Delta B=2}\, \ket{B^0_d}_{\rm SUSY}$ 
and \\
$\bra{\bar{B}^0_s}\, H_{\rm eff}^{\Delta B=2}\, \ket{B^0_s}_{\rm SUSY}$ 
given in {\tt CAUX\_H(150,151)} to the SM contributions.

\item {\tt  FDIPOLE}: Here, we  list the EDMs of  Thallium, neutron,
  Mercury, Deuteron,  Radium, and muon, as well as the  anomalous magnetic moment of
  muon, $(g_\mu-2)$.  In  the cases of the neutron  and  Mercury  EDMs,  we  present  3  and  4
  evaluations,  respectively, for the  estimation of  the theoretical
  uncertainties.
\item {\tt HiggsBoundsInputHiggsCouplingsBosons},
{\tt HiggsBoundsInputHiggsCouplingsFermions}: In these blocks,
the Higgs couplings normalized to the corresponding SM couplings
are listed
for the interface to the program {\tt HiggsBounds}~\cite{Bechtle:2008jh}.
\end{itemize}

\section{Summary}

Encouraged by  the recent observation  of a new particle  resembling a
SM-like Higgs boson at the CERN Large Hadron Collider~\cite{ATLASCMS},
in {\tt CPsuperH2.3} we have performed a number of updates to the {\tt
  CPsuperH}  code.  In  detail, we  have improved  the  computation of
CP-violating effects  on Higgs-boson  masses and mixing,  by including
stau contributions~\cite{Choi:2000wz} and  finite radiative effects on
the tau-lepton Yukawa coupling.   These effects play an important role
in  the  decays  $H_{1,2,3}\to  \gamma\gamma$,  for  large  values  of
$\tan\beta$~\cite{Carena:2011aa,LightStau}.  We  have also implemented
the  LEP  limits on  the  processes  $e^+ e^-  \to  H_i  Z, H_i  H_j$,
including the  bounds on  $H_i \to {\bar  \tau} \tau$ obtained  by CMS
with 4.6~fb$^{-1}$  of LHC data  at a centre-of-mass energy  of 7~TeV.
Finally, we have included the  decay modes of the neutral Higgs bosons
$H_{1,2,3}$  into a $Z$  boson and  a photon.   These decay  modes are
expected to become observable as more luminosity is accumulated at the
LHC.

We  have also incorporated  in {\tt  CPsuperH2.3} calculations  of the
EDMs   of  Mercury   and  $^{225}$Ra,   including  estimates   of  the
contributions due  to Schiff moments.   We have presented a  number of
numerical examples  and figures for each  of our updates,  in order to
illustrate  some typical  results obtained  by {\tt  CPsuperH2.3}.  To
enhance the synergy and  compatibility of {\tt CPsuperH2.3} with other
codes, we have created an  SLHA2 interface in accordance with the SUSY
Les  Houches Accords.   For comparison  with other  works,  we include
evaluations of the principal decay  rates and branching ratios of a SM
Higgs boson. In the Appendix, we exhibit a number of output Tables, in
order to  highlight the updates with  respect to the  older version of
{\tt CPsuperH}.

If indeed the new particle discovered by ATLAS and CMS turns out to be
a Higgs  boson, the  joint task  of theory and  experiment will  be to
establish whether it is compatible  with the SM or exhibits deviations
characteristic  of some  extension of  the SM  such as  the  MSSM. The
possibility  of CP  violation  should be  considered  in studying  the
latter possibility,  and {\tt CPsuperH2.3} is a  suitable updated tool
for this task.

\subsection*{Acknowledgements}

\noindent
We thank 
Genevieve Belanger,
Kaoru Hagiwara,
Sabine Kraml, 
Junya Nakamura and
Alexander Pukhov
for helpful discussions. 
The  work  of  JSL  is  supported   in  part  by  the  NSC  of  Taiwan
(100-2112-M-007-023-MY3), the work of JE by the London
Centre for Terauniverse Studies (LCTS), using funding from the European
Research Council via the Advanced Investigator Grant 267352,    
and    the    work    of   AP    by    the
Lancaster--Manchester--Sheffield  Consortium for  Fundamental Physics,
under STFC research grant ST/J000418/1.
Fermilab is operated by Fermi Research Alliance, LLC under Contract No.
DE-AC02-07CH11359 with the U.S. Department of Energy. Work at ANL is
supported in part by the U.S. Department of Energy under Contract No.
DE-AC02-06CH11357.

\newpage
\section*{Appendix}
Most importantly, in addition to the
{\tt SLHA2} interface,
the array for the SM parameters {\tt SMPARA\_H} has been extended to
include the SM Higgs mass,$M_{H_{\rm SM}}$, see Table \ref{tab:smpara}.
Because of these improvements, from {\tt CPsuperH2.2} to {\tt CPsuperH2.3},
the following changes to the files {\tt run} and {\tt cpsuperh2.f}
are needed:
\begin{itemize}
\item In the {\tt run} file: 
the entries {\tt SMPARA\_H(20)} and {\tt IFLAG\_H(30)} added
\item In the {\tt cpsuperh2.f} file: 
\begin{itemize}
\item{}
{\tt REAL*8 SMPARA\_H(19),SSPARA\_H(38)}
$\Longrightarrow$ 
{\tt REAL*8 SMPARA\_H(20),SSPARA\_H(38)}
\item{}
{\tt DATA NSMIN/19/}
$\Longrightarrow$ 
{\tt DATA NSMIN/20/}
\end{itemize}
\end{itemize}
Furthermore, in  the updated version,  the contents of the  array {\tt
GAMBRN(IM,IWB=2,IH)}  for the neutral  Higgs-boson decays  now include
the    corresponding    SM    branching    ratios.     To    reproduce
Figs.~\ref{fig:rpp.mstu1}  and  \ref{fig:rpp.rzp},  for  example,  one
simply needs to use the following ratios of parameter arrays:
\begin{eqnarray}
\frac{B(H_1\to\gamma\gamma)_{\rm MSSM}}
{B(H_1\to\gamma\gamma)_{\rm SM}} =
\frac{\tt GAMBRN(IM=17,IWB=3,IH=1)}
{\tt GAMBRN(IM=17,IWB=2,IH=1)}  \,;
\nonumber \\[3mm]
\frac{B(H_1\to Z\gamma)_{\rm MSSM}}
{B(H_1\to Z\gamma)_{\rm SM}} =
\frac{\tt GAMBRN(IM=19,IWB=3,IH=1)}
{\tt GAMBRN(IM=19,IWB=2,IH=1)}  \,. \nonumber
\end{eqnarray}

\def\theequation{\Alph{section}.\arabic{equation}}
\begin{appendix}

\setcounter{equation}{0}
\section{$H_i\to Z\gamma$}\label{app:h2zgamma}
For the calculation of $B(H_i\to Z\gamma)$, 
we closely follow Ref.~\cite{Djouadi:1996yq}.

The amplitude for the decay process $H_i \to Z(k_1,\epsilon_1)\
\gamma(k_2,\epsilon_2)$ can be written as
\begin{equation}
{\cal M}_{Z\gamma H_i} = -\,\frac{\alpha}{2\pi v}\left\{
S_i^{Z\gamma}(M_{H_i})\,
\left[ k_1\cdot k_2\,\epsilon_1^*\cdot\epsilon_2^*
-k_1\cdot\epsilon_2^*\,k_2\cdot\epsilon_1^* \right] \ - \
P_i^{Z\gamma}(M_{H_i})\,
\langle \epsilon_1^*\epsilon_2^* k_1 k_2\rangle
\right\}
\end{equation}
where $k_{1,2}$ are the momenta of the $Z$ boson and the photon (we note that $2k_1\cdot k_2 = M_{H_i}^2-M_Z^2$),
$\epsilon_{1,2}$ are their polarization vectors, and
$\langle \epsilon_1^*\epsilon_2^* k_1 k_2\rangle
\equiv \epsilon_{\mu\nu\alpha\beta}\epsilon_1^\mu\epsilon_2^\nu k_1^\alpha k_2^\beta$.
The decay width is given by
\begin{eqnarray}
\Gamma(H_i \to Z\gamma) &=&
\frac{\alpha G_F^2 M_W^2 s_W^2}{64\pi^4} M_{H_i}^3
\left(1-\frac{M_Z^2}{M_{H_i}^2}\right)^3 \,\left(
\left|S_i^{Z\gamma}(M_{H_i})\right|^2+
\left|P_i^{Z\gamma}(M_{H_i})\right|^2
\right)\nonumber \\
&=&
\frac{\alpha^2 M_{H_i}^3}{128\pi^3 v^2}
\left(1-\frac{M_Z^2}{M_{H_i}^2}\right)^3 \,\left(
\left|S_i^{Z\gamma}(M_{H_i})\right|^2+
\left|P_i^{Z\gamma}(M_{H_i})\right|^2
\right)\,.
\end{eqnarray}
The scalar and pseudoscalar form factors are given by
\begin{eqnarray}
S_i^{Z\gamma}(M_{H_i})\,
&=& \sum_{f=t,b,\tau} A_f^{(0)}
+A_{\widetilde\chi^\pm}^{(0)} +A_W + A_{H^\pm}
+\sum_{f=t,b,\tau} A_{\widetilde{f}} \,, \nonumber \\
P_i^{Z\gamma}(M_{H_i})\,
&=& \sum_{f=t,b,\tau} A_f^{(5)}
+A_{\widetilde\chi^\pm}^{(5)}\,,
\end{eqnarray}
with
\begin{eqnarray}
A_f^{(0),(5)} &=& 2Q_f N_C^f m_f^2\
\frac{\color{blue}I_3^f-2s_W^2 Q_f}{s_Wc_W}\ g^{S,P}_{H_{i}\bar{f}f}\ F_f^{(0),(5)}\,,
\nonumber \\
A_{\widetilde\chi^\pm}^{(0),(5)} &=&  -\ M_Z^2 \cot\theta_W
F_{\widetilde\chi^\pm}^{(0),(5)}\,,
\nonumber \\
A_W &=& M_Z^2 \cot\theta_W g_{H_{i}VV} F_W\,,
\nonumber \\
A_{H^\pm} &=& -\frac{v^2}{2c_Ws_W} g_{H_{i}H^+H^-} F_{H^\pm}\,,
\nonumber \\
A_{\widetilde{f}} &=& M_Z^2 N_C^f Q_f F_{\widetilde{f}}\,.
\end{eqnarray}
The loop functions are
\footnote{For the functions of $C_{0,2}(m^2)$,
$f(m_1,m_2,m_2)$,
$g(m_1,m_2,m_2)$, and
$C_2(m_1,m_2,m_2)$, we refer to~\cite{Djouadi:1996yq}.}
\begin{eqnarray}
F_f^{(0)}&=& C_0(m_f^2) + 4 C_2(m_f^2)\,, \nonumber \\
F_f^{(5)}&=& C_0(m_f^2) \,, \nonumber \\
F_{\widetilde\chi^\pm}^{(0)}&=&  2\sqrt{2}\
\sum_{j,k} \frac{m_{\widetilde\chi_j^\pm}}{M_W}
f\left(m_{\widetilde\chi_j^\pm},m_{\widetilde\chi_k^\pm},m_{\widetilde\chi_k^\pm}
\right) v_{Z\widetilde\chi_j^+\widetilde\chi_k^-}
g^S_{H_i\widetilde\chi_k^+\widetilde\chi_j^-}\,, \nonumber \\
F_{\widetilde\chi^\pm}^{(5)}&=&  2\sqrt{2}\ i \
\sum_{j,k} \frac{m_{\widetilde\chi_j^\pm}}{M_W}
g\left(m_{\widetilde\chi_j^\pm},m_{\widetilde\chi_k^\pm},m_{\widetilde\chi_k^\pm}
\right) v_{Z\widetilde\chi_j^+\widetilde\chi_k^-}
g^P_{H_i\widetilde\chi_k^+\widetilde\chi_j^-}\,, \nonumber \\
F_W &=& 2\left[\frac{M_{H_i}^2}{M_W^2}(1-2c_W^2)+2(1-6c_W^2)\right]
C_2(M_W^2)+4(1-4c_W^2)C_0(M_W^2)\,, \nonumber \\
F_{H^\pm} &=& 4 C_2(M_{H^\pm}^2)\,, \nonumber \\
F_{\widetilde{f}} &=& -\ \frac{4v^2}{M_Z^2 c_W s_W}
\sum_{j,k}
g_{H_i\widetilde{f}_j^*\widetilde{f}_k}
g_{Z\widetilde{f}_k^*\widetilde{f}_j}
C_2(m_{\widetilde{f}_j},m_{\widetilde{f}_k},m_{\widetilde{f}_k})\,.
\end{eqnarray}
We follow the conventions and notations of 
{\tt CPsuperH}~\cite{Lee:2003nta} for the Higgs couplings to the SM
and supersymmetric particles,
and the relevant $Z$-boson interactions are given by the 
following Lagrangian terms:
\begin{itemize}
\item\underline{$Z$-sfermion-sfermion}
\begin{eqnarray}
{\cal L}_{Z\widetilde{f}\widetilde{f}} \ = \
-ig_Z\, g_{Z\widetilde{f}_j^*\widetilde{f}_i} \
\left(\widetilde{f}_j^* \stackrel{\leftrightarrow}{\partial_\mu}
\widetilde{f}_i\right) \ Z^\mu \,,
\end{eqnarray}
where $g_Z=e/(s_Wc_W)$ and
\begin{eqnarray}
g_{Z\widetilde{f}_j^*\widetilde{f}_i} \ = \
I_3^f U_{Lj}^{\widetilde{f} *} U_{Li}^{\widetilde{f}}
-Q_f s_W^2 \delta_{ij}  \,.
\end{eqnarray}
with
$I_3^{\,u, \nu}=+1/2$ and $I_3^{\,d, e}=-1/2$.
\item\underline{$Z$-chargino-chargino}~\cite{Cheung:2011wn}
\begin{eqnarray}
{\cal L}_{Z\widetilde\chi^+\widetilde\chi^-} = -\,g_Z\,
\overline{\widetilde\chi^-_i}\,\gamma^\mu\,
\left(v_{Z\chi^+_i\widetilde\chi^-_j} - a_{Z\chi^+_i\widetilde\chi^-_j}
\gamma_5\right)\,
\widetilde\chi^-_j\,Z_\mu \, ,
\end{eqnarray}
where
\begin{eqnarray}
v_{Z\chi^+_i\widetilde\chi^-_j}& = &
\frac{1}{4}\left[
\left(C_L\right)_{i2} \left(C_L\right)_{j2}^* +
\left(C_R\right)_{i2} \left(C_R\right)_{j2}^*
\right] \ - \ c_W^2 \,\delta_{ij}\,,
\nonumber \\
a_{Z\chi^+_i\widetilde\chi^-_j} & = &
\frac{1}{4}\left[
\left(C_L\right)_{i2} \left(C_L\right)_{j2}^* -
\left(C_R\right)_{i2} \left(C_R\right)_{j2}^*
\right]\,.
\end{eqnarray}
For completeness, we recall that the $Z$-boson couplings to the quarks and leptons are given by
\begin{eqnarray}
{\cal L}_{Z\bar{f}f} = -\,g_Z\,
\bar{f}\,\gamma^\mu\,\left(v_{Z\bar{f}f} - a_{Z\bar{f}f}
\gamma_5\right)\,f\,Z_\mu \, ,
\end{eqnarray}
with $v_{Z\bar{f}f}=I_3^f/2-Q_f s_W^2$ and $a_{Z\bar{f}f}=I_3^f/2$.
\end{itemize}

%

\setcounter{equation}{0}
\section{Tables updated}\label{app:tables}
In this Appendix, we provide output Tables extended to 
include the updates implemented
since the appearance of {\tt CPsuperH2.0}.
%

%
%
%
\begin{table}[!hbp]
\caption{\label{tab:smpara}
{\it
The contents of the extended {\tt SMPARA\_H(IP)}.  }
}
\begin{center}
\begin{tabular}{|c|c|c|c|c|c|c|c|}
\hline
{\tt IP} & Parameter & {\tt IP} & Parameter
& {\tt IP} & Parameter & {\tt IP} & Parameter \\
\hline
   1 & $\alpha^{-1}_{\rm em}(M_Z)$ &
   6 & $m_\mu$                     &
  11 & $m_u (m_t^{\rm pole})$      &
  16 & $\lambda$ \\
   2 & $\alpha_s(M_Z)$             &
   7 & $m_\tau $                   &
  12 & $m_c (m_t^{\rm pole})$      &
  17 & $A$ \\
   3 & $M_Z$                       &
   8 & $m_d (m_t^{\rm pole})$      &
  13 & $m_t^{\rm pole}$            &
  18 & $\bar\rho$ \\
   4 & $\sin^2\theta_W$            &
   9 & $m_s (m_t^{\rm pole})$      &
  14 & $\Gamma_W$                  &
  19 & $\bar\eta$ \\
   5 & $m_e$                       &
  10 & $m_b (m_t^{\rm pole})$      &
  15 & $\Gamma_Z$                  &
  20 & $M_{H_{\rm SM}}$ \\
\hline
\end{tabular}
\end{center}
\end{table}
\begin{table}[!hbp]
\caption{\label{tab:sspara}
{\it
The contents of the extended {\tt SSPARA\_H(IP)}.  }
}
\begin{center}
\begin{tabular}{|c|c|c|c|c|c|c|c|c|}
\hline
  {\tt IP} & Parameter
& {\tt IP} & Parameter
& {\tt IP} & Parameter
& {\tt IP} & Parameter \\
\hline
   1 & $\tan\beta$                  &
  11 & $m_{\tilde{Q}_3}$                     &
  21 & $\Phi_{A_\tau}$            &
  31 & $|A_u|$                      \\
   2 & $M_{H^\pm}^{\rm pole}$       &
  12 & $m_{\tilde{U}_3}$                      &
  22 & $\rho_{\tilde{Q}}$                      &
  32 & $\Phi_{A_u}$                 \\
   3 & $|\mu|$                      &
  13 & $m_{\tilde{D}_3}$                     &
  23 & $\rho_{\tilde{U}}$            &
  33 & $|A_c|$                      \\
   4 & $\Phi_\mu$                   &
  14 & $m_{\tilde{L}_3}$            &
  24 & $\rho_{\tilde{D}}$            &
  34 & $\Phi_{A_c}$                 \\
   5 & $|M_1|$                      &
  15 & $m_{\tilde{E}_3}$            &
  25 & $\rho_{\tilde{L}}$            &
  35 & $|A_d|$                     \\
   6 & $\Phi_1$                     &
  16 & $|A_t|$            &
  26 & $\rho_{\tilde{E}}$            &
  36 & $\Phi_{A_d}$                     \\
   7 & $|M_2|$                      &
  17 & $\Phi_{A_t}$            &
  27 & $|A_e|$            &
  37 & $|A_s|$                     \\
   8 & $\Phi_2$                      &
  18 & $|A_b|$            &
  28 & $\Phi_{A_e}$            &
  38 & $\Phi_{A_s}$                     \\
   9 & $|M_3|$                      &
  19 & $\Phi_{A_b}$            &
  29 & $|A_\mu|$            &
  39 & ...                     \\
  10 & $\Phi_3$                      &
  20 & $|A_\tau|$            &
  30 & $\Phi_{A_\mu}$            &
  40 & ...                     \\
\hline
\end{tabular}
\end{center}
\end{table}

\begin{table}[!hbp]
\caption{\label{tab:raux_1}
{\it
The contents of the array {\tt RAUX\_H}. In {\tt RAUX\_H(22)} and {\tt RAUX\_H(23)},
the notation $h_f^0$ is for the Yukawa couplings without including the threshold
corrections. 
The notations are explained in 
Refs.~\cite{Lee:2007gn,Lee:2003nta,Carena:2000yi,Carena:2001fw,Ellis:2007kb,
Ellis:2008zy,Cheung:2009fc,Ellis:2010xm}.
For EDMs and the magnetic dipole moment (MDM) of the muon, some following
slots of the array are allocated for
the constituent contributions, see Refs.~\cite{Ellis:2008zy}, 
~\cite{Cheung:2009fc} and ~\cite{Ellis:2010xm} for details.
}}
\vspace{-0.8cm}
\begin{center}
\begin{tabular}{|cl|cl|cl|}
\hline
{\tt RAUX\_H(1)} & $\!\!\!m_b^{\rm pole}$ & 
{\tt RAUX\_H(130)} & $\!\!\!\!B(B_s\!\to\!\mu \mu)\!\times\!\! 10^7$ &
{\tt RAUX\_H(300)} &  $d_{\rm Tl}\,\,e\,$cm \\ 
{\tt RAUX\_H(2)} & $\!\!\!m_b(m_b^{\rm pole})$ & 
{\tt RAUX\_H(131)} & $\!\!\!\!B(B_d\!\to\!\tau \tau)\!\times\!\! 10^7$ &
... &  ... \\ 
{\tt RAUX\_H(3)} & $\!\!\!\alpha_s(m_b^{\rm pole})$ & 
{\tt RAUX\_H(132)} & $\!\!\!\Delta M_{B_d}^{\rm SUSY}\,{\rm ps}^{-1}$ &
{\tt RAUX\_H(310)} &  $d_n^{\rm \,CQM}\,\,e\,$cm \\ 
{\tt RAUX\_H(4)} & $\!\!\!m_c^{\rm pole}$ & 
{\tt RAUX\_H(133)} & $\!\!\!\Delta M_{B_s}^{\rm SUSY}\,{\rm ps}^{-1}$ &
... & ... \\
{\tt RAUX\_H(5)} & $\!\!\!m_c(m_c^{\rm pole})$ & 
{\tt RAUX\_H(134)} & $\!\!\!R_{B\tau\nu}$ &
{\tt RAUX\_H(320)} &  $d_n^{\rm \,PQM}\,\,e\,$cm \\ 
{\tt RAUX\_H(6)} & $\!\!\!\alpha_s(m_c^{\rm pole})$ & 
{\tt RAUX\_H(135)} & $\!\!\!\!B(B\!\to\!X_s \gamma)\!\times\!\! 10^4$ &
... & ... \\
... & ... & 
{\tt RAUX\_H(136)} & $\!\!\!\!{\cal A}_{\rm CP}(B\!\to\!X_s \gamma)\,\%$ &
{\tt RAUX\_H(330)} &  $d_n^{\rm \,QCD}\,\,e\,$cm \\ 
... & ... & 
... & ... & 
... & ... \\
... & ...  & 
{\tt RAUX\_H(200)} & $\!\!\!(d^E_e/e)\,\,$cm &
{\tt RAUX\_H(340)} &  $d_{\rm Hg}\,\,e\,$cm \\ 
{\tt RAUX\_H(10)} & $\!\!\!M_{H^\pm}^{\rm pole}$ or $M_{H^\pm}^{\rm eff.}$ & 
... & ...  & 
... & ... \\
{\tt RAUX\_H(11)} & $\!\!\!Q_t^2$  & 
{\tt RAUX\_H(210)} & $\!\!\!(d^E_u/e)\,\,$cm &
{\tt RAUX\_H(350)} &  $d_D\,\,e\,$cm \\ 
{\tt RAUX\_H(12)} & $\!\!\!Q_b^2$  & 
... & ... & 
... & ... \\
{\tt RAUX\_H(13)} & $\!\!\!Q_{tb}^2$ & 
{\tt RAUX\_H(220)} & $\!\!\!(d^E_d/e)\,\,$cm &
{\tt RAUX\_H(360)} &  $(d_\mu/e)\,\,$cm \\ 
{\tt RAUX\_H(14)} & $\!\!\!v_1(m_t^{\rm pole})$ & 
... & ... & 
... & ... \\
{\tt RAUX\_H(15)} & $\!\!\!v_1(Q_t)$ & 
{\tt RAUX\_H(230)} & $\!\!\!(d^E_s/e)\,\,$cm &
{\tt RAUX\_H(380)} &  $(a_\mu)_{\rm \,SUSY}$\\ 
{\tt RAUX\_H(16)} & $\!\!\!v_1(Q_b)$ & 
... & ... & 
... & ... \\
{\tt RAUX\_H(17)} & $\!\!\!v_1(Q_{tb})$ & 
{\tt RAUX\_H(240)} & $\!\!\!d^C_u\,\,$cm &
{\tt RAUX\_H(50)} &  $\Gamma(t\to W^+ b)$\\ 
{\tt RAUX\_H(18)} & $\!\!\!v_2(m_t^{\rm pole})$ & 
... & ... &
{\tt RAUX\_H(51)} &  $\Gamma(t\to H^+ b)$\\ 
{\tt RAUX\_H(19)} & $\!\!\!v_2(Q_t)$ & 
{\tt RAUX\_H(250)} & $\!\!\!d^C_d\,\,$cm &
{\tt RAUX\_H(52)} &  $B(t\to W^+ b)$\\ 
{\tt RAUX\_H(20)} & $\!\!\!v_2(Q_b)$ & 
... & ... &
{\tt RAUX\_H(53)} &  $B(t\to H^+ b)$\\ 
{\tt RAUX\_H(21)} & $\!\!\!v_2(Q_{tb})$ & 
{\tt RAUX\_H(260)} & $\!\!\!d^G\,\,$cm/GeV &
... & ... \\
{\tt RAUX\_H(22)} & $\!\!\!|h_t^0(m_t^{\rm pole})|$ & 
... & ... &
{\tt RAUX\_H(400)} & $\!\!\!d^C_s\,\,$cm \\
{\tt RAUX\_H(23)} & $\!\!\!|h_b^0(m_t^{\rm pole})|$ & 
{\tt RAUX\_H(270)} & $\!\!\!C_S\,\,$cm/GeV &
... & ... \\
{\tt RAUX\_H(24)} & $\!\!\!|h_t(m_t^{\rm pole})|$ & 
{\tt RAUX\_H(271)} & $\!\!\!C_P\,\,$cm/GeV &
{\tt RAUX\_H(410)} &  $d_{\rm Hg}\,\,e\,$cm\\ 
{\tt RAUX\_H(25)} & $\!\!\!|h_t(Q_t)|$ & 
{\tt RAUX\_H(272)} & $\!\!\!C_P^\prime\,\,$cm/GeV &
{\tt RAUX\_H(411)} &  $d_{\rm Hg}^{\rm I}\,\,e\,$cm\\ 
{\tt RAUX\_H(26)} & $\!\!\! |h_t(Q_{tb})|$ & 
... & ... &
{\tt RAUX\_H(412)} &  $d_{\rm Hg}^{\rm II}\,\,e\,$cm\\ 
{\tt RAUX\_H(27)} & $\!\!\! |h_b(m_t^{\rm pole})|$ & 
{\tt RAUX\_H(280)} & $\!\!\!C_{de}/m_d\,\,$cm/GeV$^2$ &
{\tt RAUX\_H(413)} &  $d_{\rm Hg}^{\rm III}\,\,e\,$cm\\ 
{\tt RAUX\_H(28)} & $\!\!\! |h_b(Q_b)|$ & 
{\tt RAUX\_H(281)} & $\!\!\!C_{se}/m_s\,\,$cm/GeV$^2$ &
{\tt RAUX\_H(414)} &  $d_{\rm Hg}^{\rm IV}\,\,e\,$cm\\ 
{\tt RAUX\_H(29)} & $\!\!\! |h_b(Q_{tb})|$ & 
{\tt RAUX\_H(282)} & $\!\!\!C_{ed}/m_d\,\,$cm/GeV$^2$ &
... & ... \\
{\tt RAUX\_H(30)} & $\!\!\! M_A^2$ & 
{\tt RAUX\_H(283)} & $\!\!\!C_{es}/m_s\,\,$cm/GeV$^2$ &
{\tt RAUX\_H(420)} &  $d_{\rm Ra}\,\,e\,$cm\\ 
{\tt RAUX\_H(31)} & $\!\!\!\!\real\widehat{\Pi}_{H^+H^-}(M_{H^\pm}^{{\rm pole}\,2})$ & 
{\tt RAUX\_H(284)} & $\!\!\!C_{eb}/m_b\,\,$cm/GeV$^2$ &
... & ... \\
%
{\tt RAUX\_H(32)} & $\!\!\!\bar\lambda_4\,v^2(m_t^{\rm pole})/2$ & 
{\tt RAUX\_H(285)} & $\!\!\!C_{ec}/m_c\,\,$cm/GeV$^2$ &
{\tt RAUX\_H(430)} &  ILEP\\ 
{\tt RAUX\_H(33)} &  $\!\!\! \bar\lambda_4(m_t^{\rm pole})$ & 
{\tt RAUX\_H(286)} & $\!\!\!C_{et}/m_t\,\,$cm/GeV$^2$ &
... & ... \\
{\tt RAUX\_H(34)} & $\!\!\! \bar\lambda_1(m_t^{\rm pole})$ & 
{\tt RAUX\_H(287)} & $\!\!\!C_{dd}/m_d\,\,$cm/GeV$^2$ &
{\tt RAUX\_H(440)} &  ${\rm ILHC7}^{H\to\tau\tau}_{4.6}$\\ 
{\tt RAUX\_H(35)} & $\!\!\! \bar\lambda_2(m_t^{\rm pole})$ & 
{\tt RAUX\_H(288)} & $\!\!\!C_{sd}/m_s\,\,$cm/GeV$^2$ &
& \\
{\tt RAUX\_H(36)} & $\!\!\! \bar\lambda_{34}(m_t^{\rm pole})$ & 
{\tt RAUX\_H(289)} & $\!\!\!C_{bd}/m_b\,\,$cm/GeV$^2$ &
& \\
... & ... & 
{\tt RAUX\_H(290)} & $\!\!\!C_{db}/m_b\,\,$cm/GeV$^2$ &
& \\
{\tt RAUX\_H(101)}&  $\!\!\! \sqrt{\hat{s}}$ & 
... & ... & 
& \\
... & ... & 
... & ... & 
& \\
\hline
\end{tabular}
\end{center}
\end{table}
\begin{table}[!hbp]
\caption{\label{tab:raux_2}
{\it
The contents of the array {\tt RAUX\_H}, continued. 
}}
\begin{center}
\begin{tabular}{|cl|cl|cl|}
\hline
{\tt RAUX\_H(501)} & $m_b(M_{H_1})$ & 
{\tt RAUX\_H(600)} & $M_{H_{\rm SM}}$ & 
... & ... \\ 
{\tt RAUX\_H(502)} & $m_t(M_{H_1})$ & 
{\tt RAUX\_H(601)} & $\Gamma(H_{\rm SM}\to ee)$ & 
{\tt RAUX\_H(651)} & $B(H_{\rm SM}\to ee)$ \\
{\tt RAUX\_H(503)} & $m_c(M_{H_1})$ & 
{\tt RAUX\_H(602)} & $\Gamma(H_{\rm SM}\to \mu\mu)$ & 
{\tt RAUX\_H(652)} & $B(H_{\rm SM}\to \mu\mu)$ \\
{\tt RAUX\_H(504)} & $m_b(M_{H_2})$ & 
{\tt RAUX\_H(603)} & $\Gamma(H_{\rm SM}\to \tau\tau)$ & 
{\tt RAUX\_H(653)} & $B(H_{\rm SM}\to \tau\tau)$ \\
{\tt RAUX\_H(505)} & $m_t(M_{H_2})$ & 
{\tt RAUX\_H(604)} & $\Gamma(H_{\rm SM}\to dd)$ & 
{\tt RAUX\_H(654)} & $B(H_{\rm SM}\to dd)$ \\
{\tt RAUX\_H(506)} & $m_c(M_{H_2})$ & 
{\tt RAUX\_H(605)} & $\Gamma(H_{\rm SM}\to ss)$ & 
{\tt RAUX\_H(655)} & $B(H_{\rm SM}\to ss)$ \\
{\tt RAUX\_H(507)} & $m_b(M_{H_3})$ & 
{\tt RAUX\_H(606)} & $\Gamma(H_{\rm SM}\to bb)$ & 
{\tt RAUX\_H(656)} & $B(H_{\rm SM}\to bb)$ \\
{\tt RAUX\_H(508)} & $m_t(M_{H_3})$ & 
{\tt RAUX\_H(607)} & $\Gamma(H_{\rm SM}\to uu)$ & 
{\tt RAUX\_H(657)} & $B(H_{\rm SM}\to uu)$ \\
{\tt RAUX\_H(509)} & $m_c(M_{H_3})$ & 
{\tt RAUX\_H(608)} & $\Gamma(H_{\rm SM}\to cc)$ & 
{\tt RAUX\_H(658)} & $B(H_{\rm SM}\to cc)$ \\
... & ... &
{\tt RAUX\_H(609)} & $\Gamma(H_{\rm SM}\to tt)$ & 
{\tt RAUX\_H(659)} & $B(H_{\rm SM}\to tt)$ \\
{\tt RAUX\_H(511)} & $M_{H_1}(\widetilde\tau\!\!\!/)$ & 
{\tt RAUX\_H(610)} & $\Gamma(H_{\rm SM}\to WW)$ & 
{\tt RAUX\_H(660)} & $B(H_{\rm SM}\to WW)$ \\
{\tt RAUX\_H(512)} & $M_{H_2}(\widetilde\tau\!\!\!/)$ & 
{\tt RAUX\_H(611)} & $\Gamma(H_{\rm SM}\to ZZ)$ & 
{\tt RAUX\_H(661)} & $B(H_{\rm SM}\to ZZ)$ \\
{\tt RAUX\_H(513)} & $M_{H_3}(\widetilde\tau\!\!\!/)$ & 
... & ... &
... & ... \\ 
... & ... &
{\tt RAUX\_H(617)} & $\Gamma(H_{\rm SM}\to \gamma\gamma)$ & 
{\tt RAUX\_H(667)} & $B(H_{\rm SM}\to \gamma\gamma)$ \\
{\tt RAUX\_H(520)} & $O_{\phi_1 1}(\widetilde\tau\!\!\!/)$ & 
{\tt RAUX\_H(618)} & $\Gamma(H_{\rm SM}\to gg)$ & 
{\tt RAUX\_H(668)} & $B(H_{\rm SM}\to gg)$ \\
{\tt RAUX\_H(521)} & $O_{\phi_1 2}(\widetilde\tau\!\!\!/)$ & 
... & ... &
... & ... \\ 
{\tt RAUX\_H(522)} & $O_{\phi_1 3}(\widetilde\tau\!\!\!/)$ & 
{\tt RAUX\_H(650)} & $\Gamma^{\rm SM}_{\rm tot}$ & 
{\tt RAUX\_H(700)} & $B^{\rm SM}_{\rm tot}=1$ \\
{\tt RAUX\_H(523)} & $O_{\phi_2 1}(\widetilde\tau\!\!\!/)$ & 
... & ... &
... & ... \\ 
{\tt RAUX\_H(524)} & $O_{\phi_2 2}(\widetilde\tau\!\!\!/)$ & 
... & ... &
... & ... \\ 
{\tt RAUX\_H(525)} & $O_{\phi_2 3}(\widetilde\tau\!\!\!/)$ & 
... & ... &
... & ... \\ 
{\tt RAUX\_H(526)} & $O_{a 1}(\widetilde\tau\!\!\!/)$ & 
... & ... &
... & ... \\ 
{\tt RAUX\_H(527)} & $O_{a 2}(\widetilde\tau\!\!\!/)$ & 
... & ... &
... & ... \\ 
{\tt RAUX\_H(528)} & $O_{a 3}(\widetilde\tau\!\!\!/)$ & 
... & ... &
... & ... \\ 
... & ... &
... & ... &
... & ... \\ 
\hline
\end{tabular}
\end{center}
\end{table}

\begin{table}[!hbp]
\caption{\label{tab:caux}
{\it
The contents of the array {\tt CAUX\_H}. 
The notations are explained in Refs.
\cite{Lee:2007gn,Lee:2003nta} and~\cite{Ellis:2007kb}.
For {\tt CAUX\_H(10,12)}, $h_q=h_q^0/(1+\Delta_q\tan\beta)$.
}}
\begin{center}
\begin{tabular}{|cl|cl|}
\hline
{\tt CAUX\_H(1)} & $h_t/|h_t|$ & 
{\tt CAUX\_H(130)} &  $S^\gamma_1(\sqrt{\hat{s}})$ \\
{\tt CAUX\_H(2)} & $h_b/|h_b|$ & 
{\tt CAUX\_H(131)} &  $P^\gamma_1(\sqrt{\hat{s}})$ \\
{\tt CAUX\_H(10)} & $\Delta_b$ & 
{\tt CAUX\_H(132)} &  $S^\gamma_2(\sqrt{\hat{s}})$ \\
{\tt CAUX\_H(11)} & $\Delta_s$ & 
{\tt CAUX\_H(133)} &  $P^\gamma_2(\sqrt{\hat{s}})$ \\
{\tt CAUX\_H(12)} & $h_s(m_t^{\rm pole})$ & 
{\tt CAUX\_H(134)} &  $S^\gamma_3(\sqrt{\hat{s}})$ \\
{\tt CAUX\_H(13)} & $h_\tau$ & 
{\tt CAUX\_H(135)} &  $P^\gamma_3(\sqrt{\hat{s}})$ \\
... & ... &
... & ... \\ 
{\tt CAUX\_H(100)} & $D^{H^0}_{1,1}({\hat{s}})$ & 
... & ... \\ 
{\tt CAUX\_H(101)} & $D^{H^0}_{1,2}({\hat{s}})$ & 
{\tt CAUX\_H(140)} & $S^g_1(\sqrt{\hat{s}})$ \\
{\tt CAUX\_H(102)} & $D^{H^0}_{1,3}({\hat{s}})$ & 
{\tt CAUX\_H(141)} & $P^g_1(\sqrt{\hat{s}})$ \\
{\tt CAUX\_H(103)} & $D^{H^0}_{1,4}({\hat{s}})$ & 
{\tt CAUX\_H(142)} & $S^g_2(\sqrt{\hat{s}})$ \\
{\tt CAUX\_H(104)} & $D^{H^0}_{2,1}({\hat{s}})$ & 
{\tt CAUX\_H(143)} & $P^g_2(\sqrt{\hat{s}})$ \\
{\tt CAUX\_H(105)} & $D^{H^0}_{2,2}({\hat{s}})$ & 
{\tt CAUX\_H(144)} & $S^g_3(\sqrt{\hat{s}})$ \\
{\tt CAUX\_H(106)} & $D^{H^0}_{2,3}({\hat{s}})$ & 
{\tt CAUX\_H(145)} & $P^g_3(\sqrt{\hat{s}})$ \\
{\tt CAUX\_H(107)} & $D^{H^0}_{2,4}({\hat{s}})$ & 
... & ... \\ 
{\tt CAUX\_H(108)} & $D^{H^0}_{3,1}({\hat{s}})$ & 
{\tt CAUX\_H(150)} & $\bra{\bar{B}^0_d}\, H_{\rm eff}^{\Delta B=2}\, \ket{B^0_d}_{\rm SUSY}$ \\
{\tt CAUX\_H(109)} & $D^{H^0}_{3,2}({\hat{s}})$ & 
{\tt CAUX\_H(151)} & $\bra{\bar{B}^0_s}\, H_{\rm eff}^{\Delta B=2}\, \ket{B^0_s}_{\rm SUSY}$ \\
{\tt CAUX\_H(110)} & $D^{H^0}_{3,3}({\hat{s}})$ & 
... & ...  \\ 
{\tt CAUX\_H(111)} & $D^{H^0}_{3,4}({\hat{s}})$ & 
{\tt CAUX\_H(160)} & $C_2^{(0)\,{\rm SM}}(m_b^{\rm pole})$  \\
{\tt CAUX\_H(112)} & $D^{H^0}_{4,1}({\hat{s}})$ &
{\tt CAUX\_H(161)} & $C_7^{(0)\,{\rm SM}}(m_b^{\rm pole})$  \\
{\tt CAUX\_H(113)} & $D^{H^0}_{4,2}({\hat{s}})$ &
{\tt CAUX\_H(162)} & $C_8^{(0)\,{\rm SM}}(m_b^{\rm pole})$  \\
{\tt CAUX\_H(114)} & $D^{H^0}_{4,3}({\hat{s}})$ &
{\tt CAUX\_H(163)} & $C_2^{(0)\,{\rm SM}+H^\pm}(m_b^{\rm pole})$  \\
{\tt CAUX\_H(115)} & $D^{H^0}_{4,4}({\hat{s}})$ &
{\tt CAUX\_H(164)} & $C_7^{(0)\,{\rm SM}+H^\pm}(m_b^{\rm pole})$  \\
{\tt CAUX\_H(116)} & $D^{H^\pm}_{H^\pm,H^\pm}({\hat{s}})$ &
{\tt CAUX\_H(165)} & $C_8^{(0)\,{\rm SM}+H^\pm}(m_b^{\rm pole})$  \\
{\tt CAUX\_H(117)} & $D^{H^\pm}_{H^\pm,G^\pm}({\hat{s}})$ &
{\tt CAUX\_H(166)} & $C_2^{(0)\,{\rm SM}+H^\pm+\tilde\chi^\pm}(m_b^{\rm pole})$  \\
{\tt CAUX\_H(118)} & $D^{H^\pm}_{G^\pm,H^\pm}({\hat{s}})$ &
{\tt CAUX\_H(167)} & $C_7^{(0)\,{\rm SM}+H^\pm+\tilde\chi^\pm}(m_b^{\rm pole})$  \\
{\tt CAUX\_H(119)} & $D^{H^\pm}_{G^\pm,G^\pm}({\hat{s}})$ &
{\tt CAUX\_H(168)} & $C_8^{(0)\,{\rm SM}+H^\pm+\tilde\chi^\pm}(m_b^{\rm pole})$  \\
... &  ... & 
{\tt CAUX\_H(169)} & $C_S(m_b^{\rm pole})$ 1/GeV  \\
... &  ... & 
{\tt CAUX\_H(170)} & $C_P(m_b^{\rm pole})$ 1/GeV \\
... &  ... & 
{\tt CAUX\_H(171)} & $C_{10}(m_b^{\rm pole})$  \\
... &  ... & 
... &  ... \\
\hline
\end{tabular}
\end{center}
\end{table}

\begin{table}[!hbp]
\caption{\label{tab:caux_2}
{\it
The contents of the array {\tt CAUX\_H}, continued. 
}}
\begin{center}
\begin{tabular}{|cl|cl|}
\hline
{\tt CAUX\_H(201)} & $\lambda_1$ & 
... & ... \\ 
{\tt CAUX\_H(202)} & $\lambda_2$ & 
... & ... \\ 
{\tt CAUX\_H(203)} & $\lambda_3$ & 
... & ... \\ 
{\tt CAUX\_H(204)} & $\lambda_4$ & 
... & ... \\ 
{\tt CAUX\_H(205)} & $\lambda_5$ & 
... & ... \\ 
{\tt CAUX\_H(206)} & $\lambda_6$ & 
... & ... \\ 
{\tt CAUX\_H(207)} & $\lambda_7$ & 
... & ... \\ 
... & ... &
... & ... \\ 
{\tt CAUX\_H(211)} & $\Delta_b\tan\beta/(1+\Delta_b \tan\beta)$ & 
... & ... \\ 
{\tt CAUX\_H(212)} & $\Delta_t\cot\beta/(1+\Delta_t \cot\beta)$ & 
... & ... \\ 
... & ... &
... & ... \\ 
{\tt CAUX\_H(221)} & $S_{\rm SM}^g(M_{H_1})$ & 
... & ... \\ 
{\tt CAUX\_H(222)} & $S_{\rm SM}^g(M_{H_2})$ & 
... & ... \\ 
{\tt CAUX\_H(223)} & $S_{\rm SM}^g(M_{H_3})$ & 
... & ... \\ 
... & ... &
... & ... \\ 
{\tt CAUX\_H(231)} & $S_{\rm SM}^\gamma(M_{H_1})$ & 
... & ... \\ 
{\tt CAUX\_H(232)} & $S_{\rm SM}^\gamma(M_{H_2})$ & 
... & ... \\ 
{\tt CAUX\_H(233)} & $S_{\rm SM}^\gamma(M_{H_3})$ & 
... & ... \\ 
... & ... &
... & ... \\ 
\hline
\end{tabular}
\end{center}
\end{table}
\begin{table}[\hbt]
\caption{\label{tab:gambrn} {\it The 
updated contents of the array {\tt
GAMBRN(IM,IWB=1,IH)} for the decays of the neutral Higgs bosons $H_{\tt IH}$.
The entry
{\tt GAMBRN(IM,IWB=1,IH)} is for
the decay width of the decay mode {\tt IM} in GeV.
And the entries
{\tt GAMBRN(IM,IWB=2,IH)} and
{\tt GAMBRN(IM,IWB=3,IH)} are the 
corresponding SM and the full SUSY branching ratios, respectively.
} }
\begin{center}
\begin{tabular}{|cl|cl|cl|}
\hline
{\tt IM} & Decay Mode& {\tt IM} & Decay Mode & {\tt IM} &  Decay Mode \\
\hline
{\tt 1}& $H_{\tt IH}\rightarrow e\bar{e}$ & {\tt 11} & $H_{\tt IH} \rightarrow Z
Z$
& .. & ...... \\
{\tt 2}& $H_{\tt IH}\rightarrow \mu\bar{\mu}$ & {\tt 12} & $H_{\tt IH}
\rightarrow H_1 Z$
& .. & ...... \\
{\tt 3}& $H_{\tt IH}\rightarrow \tau\bar{\tau}$ & {\tt 13} & $H_{\tt IH}
\rightarrow H_2 Z$
& .. &  ...... \\
{\tt 4}& $H_{\tt IH}\rightarrow d\bar{d}$ & {\tt 14} & $H_{\tt IH} \rightarrow
H_1 H_1$
& .. &  ...... \\
{\tt 5}& $H_{\tt IH}\rightarrow s\bar{s}$ & {\tt 15} & $H_{\tt IH} \rightarrow
H_1 H_2$
& .. &  ...... \\
{\tt 6}& $H_{\tt IH}\rightarrow b\bar{b}$ & {\tt 16} & $H_{\tt IH} \rightarrow
H_2 H_2$
& .. &  ...... \\
{\tt 7}& $H_{\tt IH}\rightarrow u\bar{u}$ & {\tt 17} & $H_{\tt IH} \rightarrow
\gamma\gamma$
& .. &  ...... \\
{\tt 8}& $H_{\tt IH}\rightarrow c\bar{c}$ & {\tt 18} & $H_{\tt IH} \rightarrow
g\,g$
& .. &  ...... \\
{\tt 9}& $H_{\tt IH}\rightarrow t\bar{t}$ & {\color{red}\tt 19} & 
{\color{red}$H_{\tt IH} \rightarrow Z\,\gamma$}
& .. &  ...... \\
{\tt 10}& $H_{\tt IH}\rightarrow WW$ & .. & ......
& {\tt ISMN} &  {\tt GAMSM} \\
\hline
{\tt IM} & Decay Mode& {\tt IM} & Decay Mode & {\tt IM} &  Decay Mode \\
\hline
{\tt ISMN+1}&$H_{\tt IH}\rightarrow \widetilde{\chi}^0_1\widetilde{\chi}_1^0$&
{\tt ISMN+11}& $H_{\tt IH}\rightarrow \widetilde{\chi}^+_1\widetilde{\chi}_1^-$
& {\tt ISMN+21} &  $H_{\tt IH}\rightarrow \widetilde{b}^*_2\widetilde{b}_1$\\
{\tt ISMN+2}&$H_{\tt IH}\rightarrow \widetilde{\chi}^0_1\widetilde{\chi}_2^0$&
{\tt ISMN+12} & $H_{\tt IH}\rightarrow \widetilde{\chi}^+_1\widetilde{\chi}_2^-$
& {\tt ISMN+22} &  $H_{\tt IH}\rightarrow \widetilde{b}^*_2\widetilde{b}_2$\\
{\tt ISMN+3}&$H_{\tt IH}\rightarrow \widetilde{\chi}^0_1\widetilde{\chi}_3^0$&
{\tt ISMN+13} & $H_{\tt IH}\rightarrow \widetilde{\chi}^+_2\widetilde{\chi}_1^-$
& {\tt ISMN+23} &  $H_{\tt IH}\rightarrow
\widetilde{\tau}^*_1\widetilde{\tau}_1$\\
{\tt ISMN+4}&$H_{\tt IH}\rightarrow \widetilde{\chi}^0_1\widetilde{\chi}_4^0$&
{\tt ISMN+14} & $H_{\tt IH}\rightarrow \widetilde{\chi}^+_2\widetilde{\chi}_2^-$
& {\tt ISMN+24} &  $H_{\tt IH}\rightarrow
\widetilde{\tau}^*_1\widetilde{\tau}_2$\\
{\tt ISMN+5}&$H_{\tt IH}\rightarrow \widetilde{\chi}^0_2\widetilde{\chi}_2^0$&
{\tt ISMN+15} & $H_{\tt IH}\rightarrow \widetilde{t}^*_1\widetilde{t}_1$
& {\tt ISMN+25} &  $H_{\tt IH}\rightarrow
\widetilde{\tau}^*_2\widetilde{\tau}_1$\\
{\tt ISMN+6}&$H_{\tt IH}\rightarrow \widetilde{\chi}^0_2\widetilde{\chi}_3^0$&
{\tt ISMN+16} & $H_{\tt IH}\rightarrow \widetilde{t}^*_1\widetilde{t}_2$
& {\tt ISMN+26} &  $H_{\tt IH}\rightarrow
\widetilde{\tau}^*_2\widetilde{\tau}_2$\\
{\tt ISMN+7}&$H_{\tt IH}\rightarrow \widetilde{\chi}^0_2\widetilde{\chi}_4^0$&
{\tt ISMN+17} & $H_{\tt IH}\rightarrow \widetilde{t}^*_2\widetilde{t}_1$
& {\tt ISMN+27} &  $H_{\tt IH}\rightarrow
\widetilde{\nu}^*_\tau\widetilde{\nu}_\tau$\\
{\tt ISMN+8}&$H_{\tt IH}\rightarrow \widetilde{\chi}^0_3\widetilde{\chi}_3^0$&
{\tt ISMN+18} & $H_{\tt IH}\rightarrow \widetilde{t}^*_2\widetilde{t}_2$
& .. &  ...... \\
{\tt ISMN+9}&$H_{\tt IH}\rightarrow \widetilde{\chi}^0_3\widetilde{\chi}_4^0$&
{\tt ISMN+19} & $H_{\tt IH}\rightarrow \widetilde{b}^*_1\widetilde{b}_1$
& {\tt ISMN+ISUSYN} &  {\tt GAMSUSY} \\
{\tt ISMN+10}&$H_{\tt IH}\rightarrow \widetilde{\chi}^0_4\widetilde{\chi}_4^0$&
{\tt ISMN+20} & $H_{\tt IH}\rightarrow \widetilde{b}^*_1\widetilde{b}_2$
& {\tt ISMN+ISUSYN+1} &  {\tt GAMSM+GAMSUSY} \\
\hline
\end{tabular}
\end{center}
\end{table}

\end{appendix}

\end{document}